\documentclass[useAMS,usenatbib]{mn2e}

\usepackage{aecompl}
\usepackage[toc,page]{appendix}

\usepackage{graphicx}
\usepackage{amssymb}
\usepackage{amsmath}
\usepackage{aas_macros}
\usepackage{deluxetable}
\usepackage{lscape}
\usepackage{rotating}
\usepackage[symbol*]{footmisc}
\usepackage{gensymb}
\usepackage{tikz}
\def\checkmark{\tikz\fill[scale=0.4](0,.35) -- (.25,0) -- (1,.7) -- (.25,.15) -- cycle;}
\def\cross{\tikz\draw[scale=0.25](0,1) -- (0.8,0) (0,0) -- (0.8,1);}



\title[Filament fragmentation length-scales]{Determining the presence of characteristic fragmentation length-scales in filaments}
\author[S. D. Clarke et al.]{S. D. Clarke$^{1}$\thanks{E-mail: clark@ph1.uni-koeln.de }, G. M. Williams$^{2}$, J. C. Ib\'a\~nez-Mej\'{\i}a$^{1}$ and S. Walch$^{1,3}$. \\$^{1}$I. Physikalisches Institut, Universit{\"a}t zu K{\"o}ln, Z{\"u}lpicher Str. 77, D-50937 K{\"o}ln, Germany \\$^{2}$Centre for Astrophysics Research, School of Physics, Astronomy and Mathematics, University of Hertfordshire,\\ College Lane, Hatfield, Al10 9AB, UK \\$^{3}$Cologne Centre for Data and Simulation Science, Univeristy of Cologne, Cologne, Germany,\\ www.cds.uni-koeln.de}

\newcommand{\OO}{_{_{\rm O}}}

\begin{document}

\date{}

\pagerange{\pageref{firstpage}--\pageref{lastpage}} \pubyear{2002}

\maketitle

\label{firstpage}

\begin{abstract}
Theories suggest that filament fragmentation should occur on a characteristic fragmentation length-scale. This fragmentation length-scale can be related to filament properties, such as the width and the dynamical state of the filament. Here we present a study of a number of fragmentation analysis techniques applied to filaments, and their sensitivity to characteristic fragmentation length-scales. We test the sensitivity to both single-tier and two-tier fragmentation, i.e. when the fragmentation can be characterised with one or two fragmentation length-scales respectively. The nearest neighbour separation, minimum spanning tree separation and two-point correlation function are all able to robustly detect characteristic fragmentation length-scales. The Fourier power spectrum and the N$^{th}$ nearest neighbour technique are both poor techniques, and require very little scatter in the core spacings for the characteristic length-scale to be successfully determined. We develop a null hypothesis test to compare the results of the nearest neighbour and minimum spanning tree separation distribution with randomly placed cores. We show that a larger number of cores is necessary to successfully reject the null hypothesis if the underlying fragmentation is two-tier, $N \gtrsim 20$. Once the null is rejected we show how one may decide if the observed fragmentation is best described by single-tier or two-tier fragmentation, using either Akaike's information criterion or the Bayes factor. The analysis techniques, null hypothesis tests, and model selection approaches are all included in a new open-source \textsc{Python/C} library called \textsc{FragMent}.

\end{abstract}


\section{Introduction}\label{SEC:INTRO}%
The Herschel Space observatory has shown that filaments are abundant in the ISM, both in the diffuse and the dense molecular gas. Moreover, filaments are observed to fragment, harbouring numerous protostellar and prestellar cores, acting as an intermediate step between molecular clouds and star formation \citep{And10, Kon15, Mar16}. 

Due to their prevalence and importance in star formation, there have been a number of theoretical papers studying the properties of filaments. It has been shown that a filamentary geometry leads to longer global collapse timescales compared to equally dense spheres. As such, smaller perturbations are able to grow and produce cores before global collapse over takes them, making filaments excellent sites for core formation \citep{BurHar04,Pon12,Cla15}. Furthermore, perturbation studies suggest that there exist a characteristic fragmentation scale. Such studies of equilibrium filaments show that the fastest growing perturbation has a wavelength of roughly four times the diameter \citep{InuMiy92, FisMar12}. Perturbations with this fastest growing wavelength dominate over others when realistic density perturbations are used, leading to a filament fragmented into a series of roughly evenly spaced cores \citep{InuMiy97,Cla16}. 

Recent work that goes beyond the equilibrium model shows that fragmentation is more complicated; the fastest growing mode is no longer linked to the diameter but rather to the accretion rate onto, and temperature of the filament \citep{Cla16}. The addition of turbulence further complicates the fragmentation process. When gravity dominates over turbulence a two-tier hierarchical fragmentation occurs; the filament first fragments with a large-scale separation determined by the fastest growing mode; then the newly formed clumps subsequently fragment on the Jeans length. When turbulence dominates, the fragmentation of the filament is no longer linked to the filament's properties but to the turbulent properties, namely the mode with the most energy \citep{Cla17}. A high level of turbulence also leads to filaments fragmenting into sub-filaments. Cores are then located on these sub-filaments or at the junction of two or more sub-filaments \citep{Cla17, Cla18}. This behaviour has been seen in recent high resolution observations \citep{TafHac15,Hac17,Hac18}.

As the spacing of cores may contain information about the filament and its turbulence, observers have attempted to measure it \citep{Zha09,Jac10,Mie12,Bus13,Kai13,Lu14,Beu15,Hen16,Tei16,Kai17,Wil18}. A variety of techniques, as detailed in the next section, have been used to determine if a characteristic fragmentation length-scale exists. However, it is not clear how sensitive these techniques are when applied to filaments, or when multiple fragmentation length-scales are present. 

In this paper, we present a study of the sensitivities of fragmentation analysis techniques commonly used in the literature. In Section \ref{SEC:TECH}, we detail the fragmentation analysis techniques studied. In Section \ref{SEC:SYNFIL}, we show the method used to produce synthetic filaments with one or two characteristic fragmentation length-scales. In Section \ref{SEC:RES}, we present the results from applying the fragmentation analysis techniques to these synthetic filaments. In Section \ref{SEC:DIS}, we develop methods for testing the statistical significance of the results and determining the number of characteristic fragmentation length-scales. In Section \ref{SEC:CON}, we summarise our conclusions. 

\section{Fragmentation techniques}\label{SEC:TECH}%

The fragmentation analysis techniques used in the literature can be split into two groups: those that work on point data (i.e. core locations), and those that work on column density maps/profiles. Here we investigate five different techniques, four point data techniques: the \textit{nearest neighbour separation}, the \textit{minimum spanning tree}, the \textit{N$^{th}$ nearest neighbour separation}, and the \textit{two-point correlation function}; and one column density technique: the \textit{Fourier power spectrum}. We compile these five methods into an open-source \textsc{Python/C} package called \textsc{FragMent} for use by the community. 

\subsection{Nearest neighbour separation}%

The nearest neighbour (NN) separation is a simple method for studying the fragmentation in filaments and to determine the existence of characteristic length-scales. This technique requires one to find the nearest neighbour of every core, and then study the mean and standard deviation of this distribution. If the spread around the average is small then it is indicative of a characteristic fragmentation length-scale. Typically, to test the statistical significance of the resulting distribution of spacings, one compares to a suite of Monte Carlo simulations of randomly placed cores with no characteristic spacing. 

\subsection{Minimum spanning tree}%

The minimum spanning tree (MST) is a subset of the complete graph that minimises the sum of the edge lengths while connecting every data point without loops \citep{GowRos69}. For a straight filament, the minimum spanning tree reduces to a set of edges which join a core to the core that is next along the filament. The unique distribution of edge lengths is examined for characteristic length-scales in much the same way as the nearest neighbour separation distribution. Comparisons with minimum spanning trees constructed using randomly placed cores can be used to estimate the statistical significance of the results.

\subsection{N$^{th}$ nearest neighbour separation}%

The N$^{th}$ nearest neighbour separation is an extension of the nearest neighbour technique. Here the separation between every pair of cores is measured. This results in a distribution of spacings for each N$^{th}$ neighbour. The mean and standard deviation of each of these distributions is plotted against each other. If there exists a local minimum in this plot then it is considered a characteristic length-scale. This method has the advantage that it is not dependent on only the nearest neighbour separation, but contains information on multiple scales. 

\subsection{Two-point correlation function} \label{SSEC:2point}%

The two-point correlation function at a distance $r$ is defined as \citep{LanSza93}:
\begin{equation}
W(r) = \frac{DD(r) - 2DR(r) + RR(r)}{RR(r)},
\end{equation}
where $DD(r)$ is the normalised distribution of separation distances of the complete graph constructed from the data points (i.e. core locations); $RR(r)$ is the same but for the complete graph constructed from a set of randomly placed positions of the same size as the data points; $DR(r)$ is the normalised distribution of separation distances between every data point and every random position. Due to the use of random positions, this calculation is repeated tens of thousands of times so that the distribution of $W$ is well sampled. If the data points are randomly distributed then $W(r) \, \sim \, 0$. Distances where $W(r) > 0$ correspond to enhancement and suggest a characteristic length-scale; distances where $W(r) < 0$ are those that exhibit a deficit of separations compared to a random distribution.

\subsection{Fourier power spectrum}%

A Fourier transform is able to detect the presence of periodic features, making it useful to study the existence of characteristic length-scales. There are two possible methods to construct a Fourier transform to study fragmentation scales: a 1D Fourier transform of the column density profile along the spine of the filament, and the 2D Fourier transform of the entire column density map. The information of the 2D Fourier transform is compressed by constructing a radially averaged power spectrum. Here we only investigate the 1D Fourier transform as it is the most sensitive. A characteristic length-scale will appear in a power spectrum as an excess at a certain wavelength.

\section{Synthetic filaments}\label{SEC:SYNFIL}%

\begin{figure}
\includegraphics[width = 0.95\linewidth]{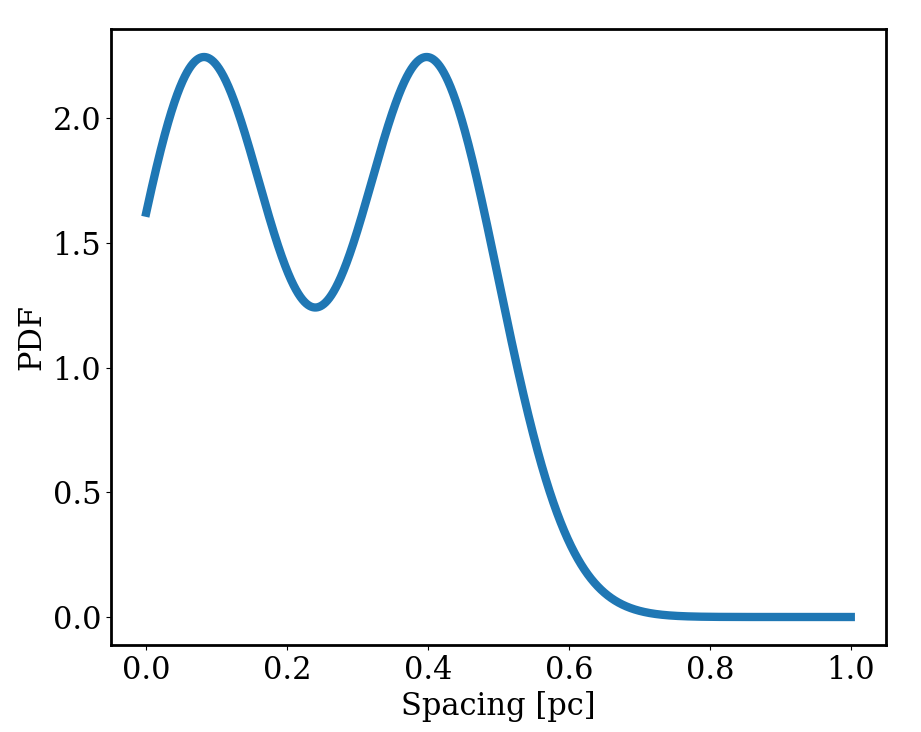}
\caption{An example probability density function for the spacing between cores when one considers two-tier fragmentation. The parameters are: $\mu \, = 0.4 \, \rm pc$, $\sigma \, = \, 0.1 \, \rm pc$ and $a \, = \, 5$. Thus the small-scale spacing is 0.08 pc. See 3$^{rd}$ panel of figure \ref{fig::synfil}}
\label{fig::pdf}
\end{figure} 

To test these techniques, one requires column density maps of synthetic filaments with known characteristic fragmentation length-scales. A synthetic filament is made of two components: the filament itself, with column density $\Sigma_f$, and the cores, with column density $\Sigma_c$. The filament can be modelled as a Plummer-like profile \citep{WhiWar01}, and the cores can be modelled as 2D Gaussians. 

A Plummer-like column density profile takes the form:
\begin{equation}
\Sigma_{f}(x,y) = \frac{\Sigma_p}{\left[ 1 + ( x/r_{_{\rm flat}} )^2 \right]^{\frac{p-1}{2}}},
\end{equation}
where $\Sigma_p$ is the central column density of the filament, $r_{flat}$ is the inner flattening radius, and $p$ is the exponent defining the power-law at large radii. Observations suggest that filaments may be characterised by Plummer-like profiles with $p \sim  2$ \citep{Arz11}.

The 2D Gaussian profile takes the form:
\begin{equation}
\Sigma_{c}(x,y,x\OO,y\OO) = \Sigma_g e^{-r^2/(2w^2)},
\end{equation}
where $r\equiv\sqrt{(x-x\OO)^2 + (y-y\OO)^2}$, $(x\OO,y\OO)$ are the co-ordinates of the core, $\Sigma_g$ is the central column density of the core, and $w$ is related to the Full Width Half Maximum (FWHM) by the expression $ \sqrt{8\ln{2}} w $. Note that these cores are perfectly circular and have the same central column density; this is not the case for real filaments but will not affect the point data techniques as they require only the core centre and is the optimal case for the Fourier transform. 

The synthetic filament is described by the combination of these two components:
\begin{equation}
\Sigma_{_{\rm tot}}(x,y) = \Sigma_f(x,y) + \sum_i^N{\Sigma_{c}(x,y,x_{i},y_{i})},
\end{equation}
where $N$ is the number of cores in the filament. Thus the model contains 5 parameters: $\Sigma_p, r_{_{\rm flat}}, p, \Sigma_g, w $. The values of these parameters, unless otherwise stated, are: 
\begin{align*}
\Sigma_p =& \, 10^{22} \, \rm{cm^{-2}}, \\
r_{_{\rm flat}} =& \, 0.05 \, \rm{pc}, \\
p =& \, 2.0, \\
\Sigma_g =& \,2 \, \times \, 10^{22} \, \rm{cm^{-2}}, \\
w =& \, 0.02 \, \rm{pc}.
\end{align*}
The column density map is $4 \rm{pc} \times 4 \rm{pc}$.

\begin{figure*}
\includegraphics[width=0.9\linewidth]{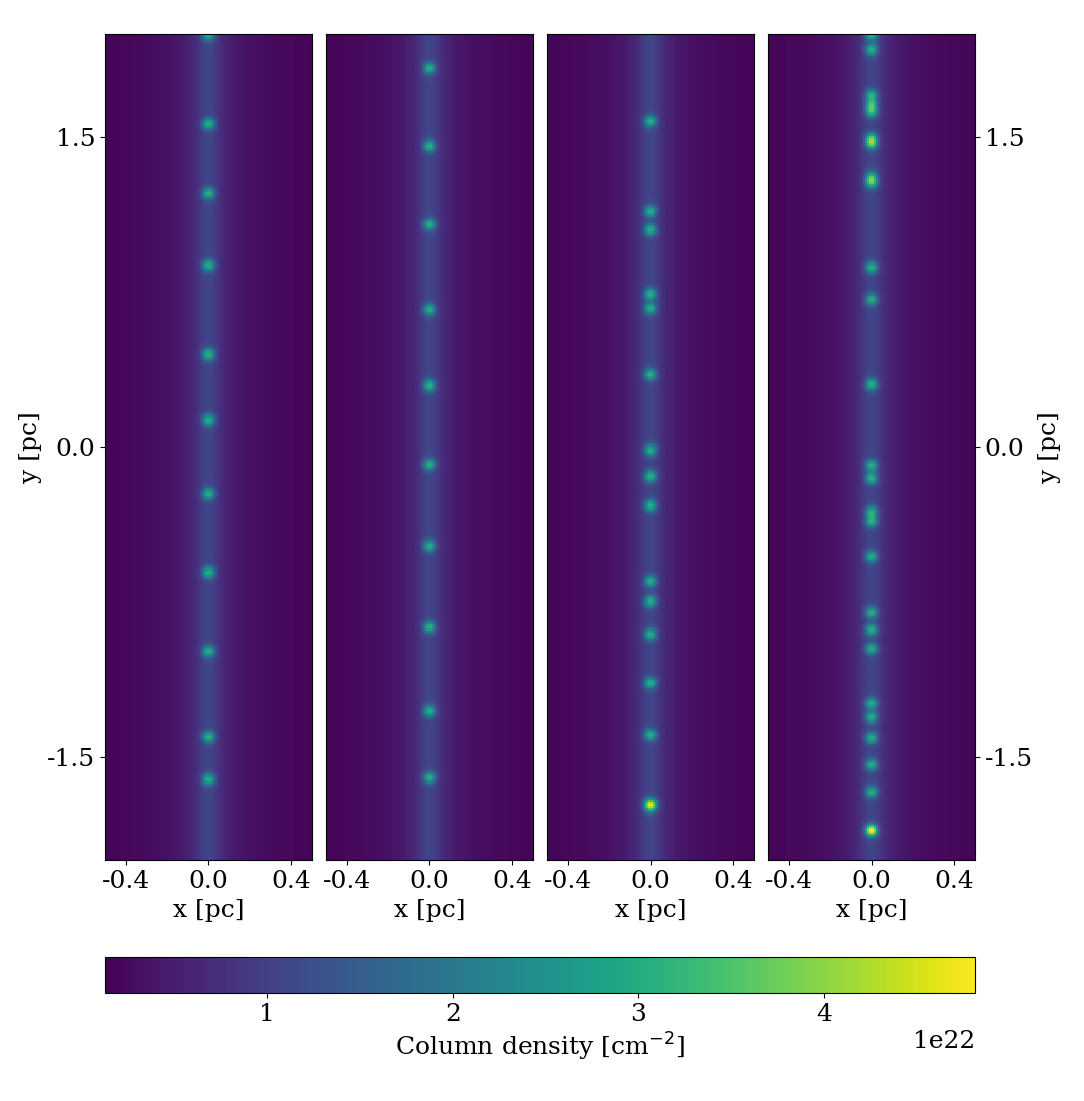}
\caption{Four examples of synthetic column density plots. (First panel) A filament with a single characteristic core spacing of 0.4 pc and a standard deviation 0.1 pc. (Second panel) A filament with a single characteristic core spacing of 0.4 pc and a standard deviation 0.04 pc. (Third panel) A filament with two characteristic core spacings, 0.08 pc and 0.4 pc, with a standard deviation around both of 0.1 pc. (Fourth panel) A filament with two characteristic core spacings, 0.08 pc and 0.4 pc, with a standard deviation around both of 0.04 pc.}
\label{fig::synfil}
\end{figure*}

The spacing between two cores is determined by sampling a random number from a Gaussian distribution. The mean of the Gaussian, $\mu$, is equivalent to a characteristic fragmentation length-scale, and the standard deviation, $\sigma$, is equivalent to some underlying scatter around this length-scale. Cores continue to be placed until the final position is greater than 4 pc, the extent of the map. The cores are placed on the long axis of the filament. For filaments mimicking a two-tier fragmentation process, the spacing is a random number drawn from a probability density function made up of two Gaussian distributions (see figure \ref{fig::pdf}). For simplicity, the mean of the Gaussian approximating the small-scale spacing is taken to be some factor, $a$, of the mean of the Gaussian approximating the large-scale spacing. The standard deviation of the small-scale spacing is the same as that of the large-spacing. Figure \ref{fig::synfil} shows four examples of synthetic filaments with cores along them.

For a fragmentation analysis technique to be successful it should be able to determine the mean of the Gaussian distribution(s) from which the core spacing was sampled. The locations of the cores are determined by using a dendrogram\footnote{We use the {\sc astrodendro} Python package, http://www.dendrograms.org}. The parameters used for the dendrogram are: \textsc{min$\_$value}=$10^{22} \rm{cm^{-2}}$, \textsc{min$\_$delta}=10$^{22} \rm{cm^{-2}}$, \textsc{min$\_$npix}=4. This is done to better approximate the limitations of observations, e.g. the introduction of a minimum core separation due to the beam-size and core identification method. In our model the choice of \textsc{min$\_$npix} leads to a minimum core separation of 0.04 pc. This causes one or two core separations to be missed in some realisations.

These synthetic filaments are perfectly straight and aligned with the $y$-axis. This is not the case for real filaments. However, one may first `straighten' the filament before beginning the analysis, i.e. identifying the spine of the filament, straighten it and align it with the $y$-axis \citep[referred to as a deprojected filament by][Watkins et al. in prep.]{Wil18}. We encourage this technique, where possible, as it effectively reduces the problem to 1D and removes the complications of filament curvature. An algorithm for straightening filaments is provided in \textsc{FragMent}, requiring the column density / integrated intensity map and the spine of the filament (see appendix \ref{app::straight} for details).

\begin{figure*}
\includegraphics[width=0.45\linewidth]{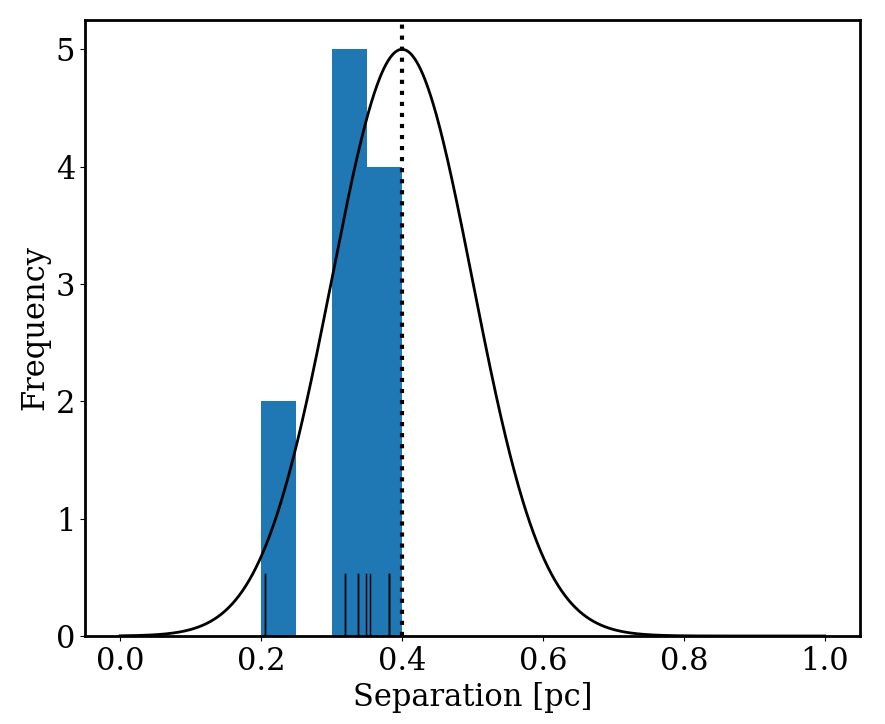}
\includegraphics[width=0.45\linewidth]{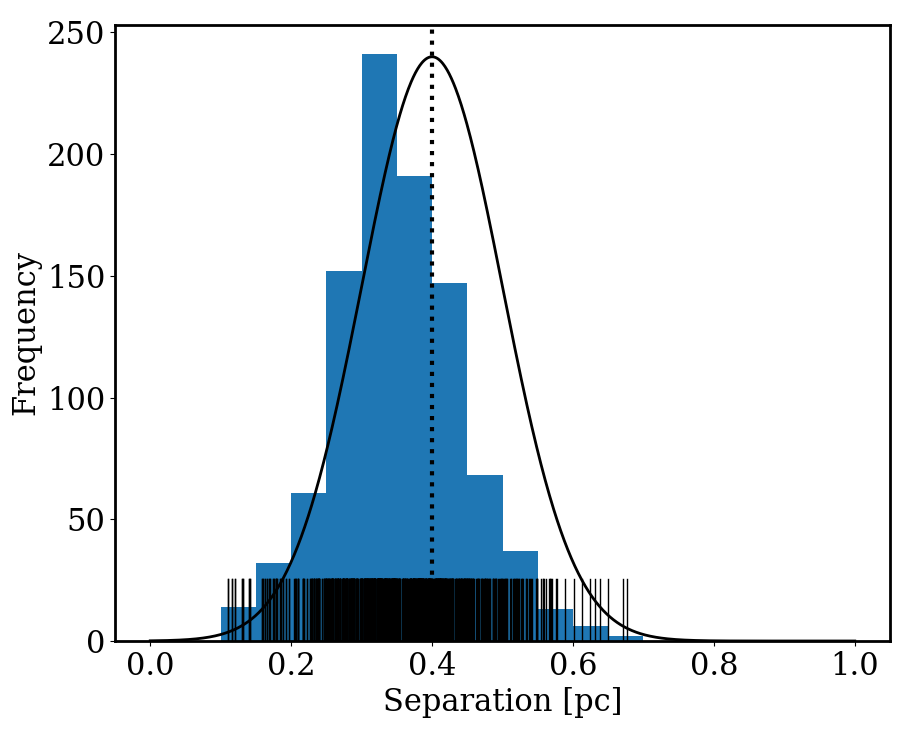}
\includegraphics[width=0.45\linewidth]{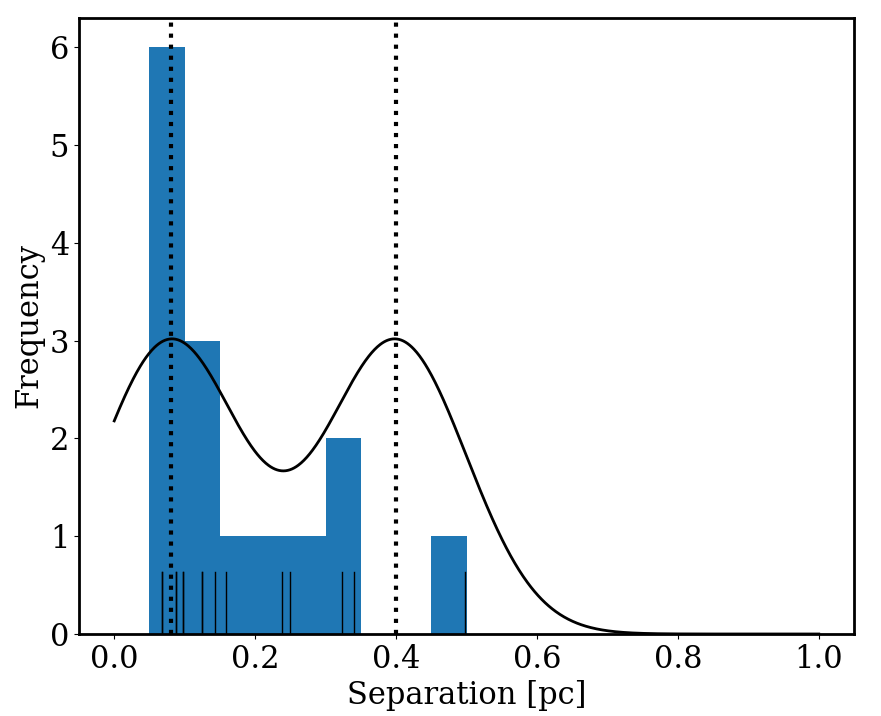}
\includegraphics[width=0.45\linewidth]{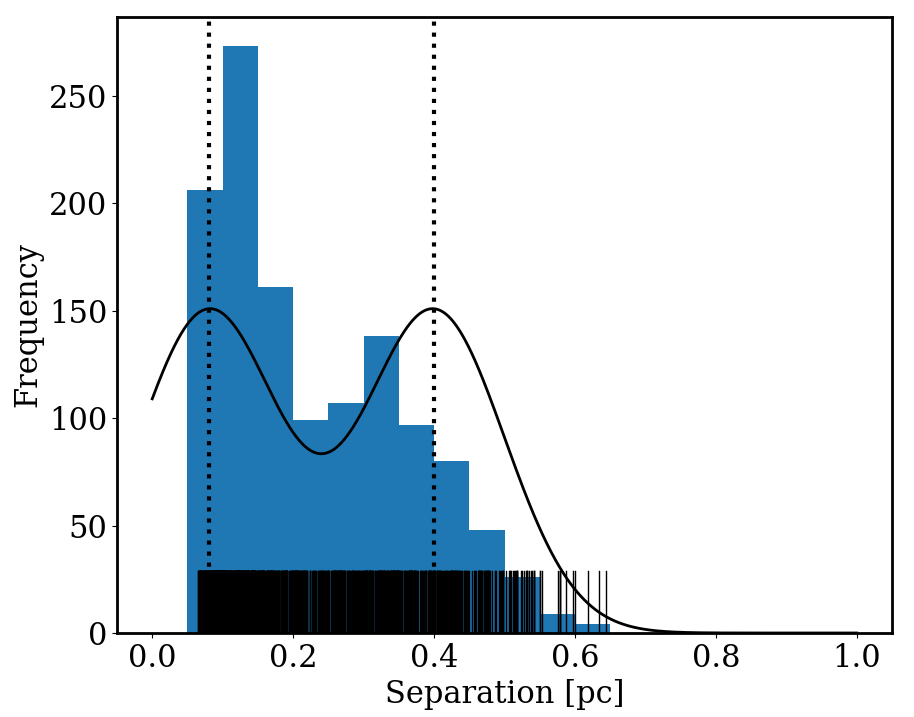}
\caption{(Left column) The nearest neighbour separation distribution from one realisation using the fiducial fragmentation parameters. (Right column) The nearest neighbour separation distribution when considering the separations from all 100 realisations of the fiducial fragmentation parameters. (Top row) When the fragmentation is single-tier. (Bottom row) When the fragmentation is two-tier. The vertical solid black lines show the individual separations. The black dotted line shows the characteristic spacings used to construct the filament, 0.4 pc for the large-scale spacing and 0.08 pc for the small-scale spacing in the two-tier case. Overlaid is the underlying PDF from which the core separations were sampled.}
\label{fig::NNS_dist}
\end{figure*} 

\section{Results}\label{SEC:RES}%

Here we present the results of each technique separately. For a set of parameters describing a synthetic filament we produce 100 realisations to achieve good number statistics. The fiducial parameters for single tier fragmentation are: $\mu \, = 0.4 \, \rm pc$, $\sigma \, = \, 0.1 \, \rm pc$. The fiducial parameters for two-tier fragmentation are: $\mu \, = 0.4 \, \rm pc$, $\sigma \, = \, 0.1 \, \rm pc$, and $a \, = \, 5$. For the fiducial single-tier fragmentation parameters the 100 realisations contain between 7 and 11 cores. For the fiducial two-tier fragmentation parameters the 100 realisations contain between 11 and 16 cores. 

When a single realisation is shown, it is always the same realisation (seed 1) so that the methods can be compared using the same data. The results from all 100 realisations are combined and shown, where possible, to study the method's sensitivity when one has robust number statistics and sample size is not an issue. A summary of the positives and negatives of each method is shown in table \ref{tab::Pos_Neg_methods}.

\subsection{Nearest neighbour separation}\label{SSEC:RES:NNS}%

The nearest neighbour separation for $N$ cores returns a distribution of $N$ separations. The top row of figure \ref{fig::NNS_dist} shows the distribution of separations from one realisation of the fiducial single-tier fragmentation (left), and the distribution from all 100 realisations (right). Both distributions are peaked and relatively narrow, indicating that the method is detecting the underlying characteristic separation. 

The median, $\mathcal{M}$, and mean, $M$, of the single realisation are 0.338 pc and 0.325 pc respectively, with an interquartile range, $\mathcal{I}$, and standard deviation, $S$, of 0.049 pc and 0.060 pc. The average of the distribution is lower than that of the characteristic separation, 0.4 pc, and the width of the distribution is also narrower than that of the probability density function the spacing was sampled from, 0.1 pc. For the distribution of separations from all 100 realisations: $\mathcal{M}=$0.347 pc with $\mathcal{I}$=0.117 pc, and $M$=0.353 pc with $S$=0.091 pc. Thus, the nearest neighbour separation technique systematically leads to lower estimates of the characteristic spacing, if one exists.

The bottom row of figure \ref{fig::NNS_dist} shows the distribution of separations from one realisation of the fiducial two-tier fragmentation (left), and the distribution from all 100 realisations (right). The distribution from the single realisation is strongly peaked at $\sim$ 0.08 pc, but does not appear to be bimodal. The distribution from all 100 realisations is clearly bimodal, with peaks at $\sim$ 0.1 pc and $\sim$ 0.35 pc; however, the peak at small scales is significantly larger than that at larger scales, which is not the case in the underlying PDF of spacings used. For the single realisation distribution: $\mathcal{M}$=0.124 pc with $\mathcal{I}$=0.151 pc, and $M$=0.180 pc with $S$=0.121 pc. For all 100 realisations: $\mathcal{M}=$0.194 pc with $\mathcal{I}$=0.216 pc, and $M$=0.232 pc with $S$=0.131 pc. It is clear that the median/mean and interquartile range/standard deviation are poor descriptors of a bimodal distribution. Table \ref{tab::NNS_MST_RES} summaries the averages and width measurements.  

In summary, the nearest neighbour separation distribution is a simple method which can detect the presence of a single characteristic fragmentation length-scale. However, by definition it selects only the smallest separation of each core, leading to a bias to underestimate the characteristic fragmentation length-scale. This sensitivity to smaller separations also causes the method to be weak at detecting multiple fragmentation length-scales.

\begin{table*}
\centering
\begin{tabular}{@{}*7l@{}}
\hline\hline
Method                  & Fragmentation type & Realisations & Median $\mathcal{M}$ & Interquartile range $\mathcal{I}$ & Mean $M$   & Standard deviation $S$  \\ \hline
Nearest neighbour       & Single-tier        & Single       & 0.338 pc             & 0.049 pc                          & 0.325 pc   & 0.060 pc \\
Nearest neighbour       & Single-tier        & All 100      & 0.347 pc             & 0.117 pc                          & 0.353 pc   & 0.091 pc \\
Nearest neighbour       & Two-tier           & Single       & 0.124 pc             & 0.151 pc                          & 0.180 pc   & 0.121 pc \\
Nearest neighbour       & Two-tier           & All 100      & 0.194 pc             & 0.216 pc                          & 0.232 pc   & 0.131 pc \\ \hline
Minimum spanning tree   & Single-tier        & Single       & 0.367 pc             & 0.041 pc                          & 0.355 pc   & 0.059 pc \\
Minimum spanning tree   & Single-tier        & All 100      & 0.395 pc             & 0.129 pc                          & 0.396 pc   & 0.098 pc \\
Minimum spanning tree   & Two-tier           & Single       & 0.244 pc             & 0.207 pc                          & 0.241 pc   & 0.126 pc \\
Minimum spanning tree   & Two-tier           & All 100      & 0.322 pc             & 0.270 pc                          & 0.307 pc   & 0.153 pc \\ \hline
\end{tabular}
\centering
\caption{The average and widths of the separation distributions resulting from applying the nearest neighbour and minimum spanning tree techniques to the single-tier and two-tier fragmentation synthetic filaments.}
\label{tab::NNS_MST_RES}
\end{table*}

\begin{figure*}
\includegraphics[width=0.45\linewidth]{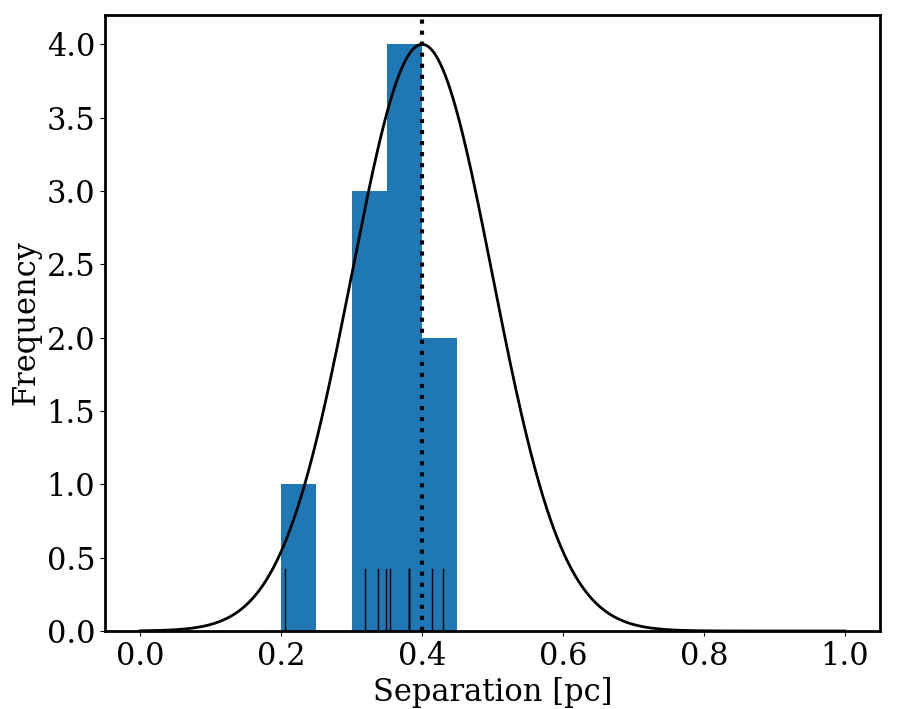}
\includegraphics[width=0.45\linewidth]{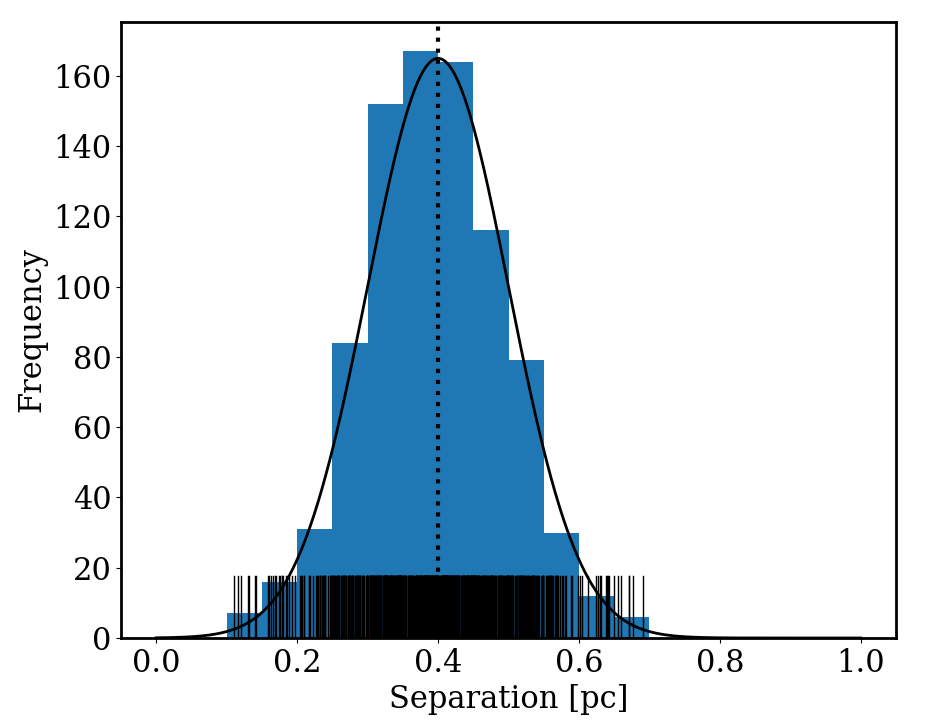}
\includegraphics[width=0.45\linewidth]{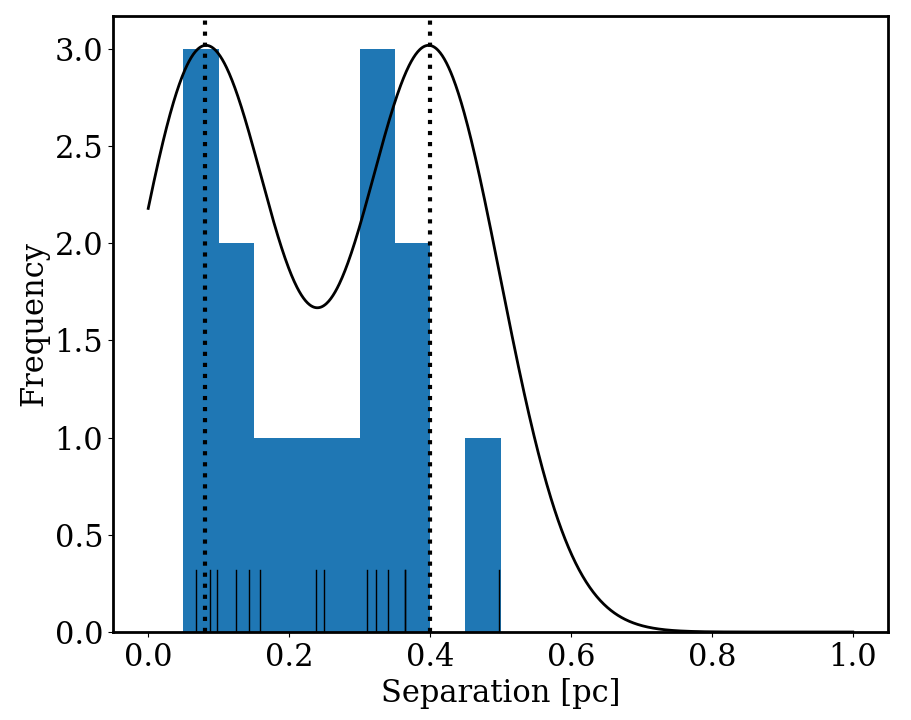}
\includegraphics[width=0.45\linewidth]{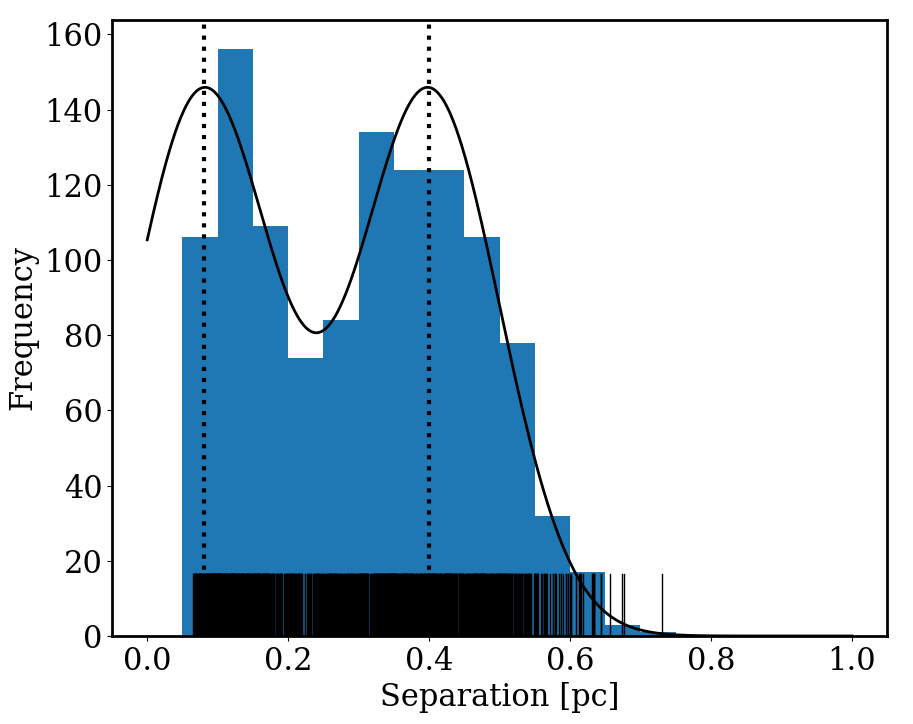}
\caption{The same as figure \ref{fig::NNS_dist} for the minimum spanning tree method.}
\label{fig::MST_dist}
\end{figure*} 

\subsection{Minimum spanning tree}\label{SSEC:RES:MST}%

The minimum spanning tree for $N$ cores returns $N-1$ edges. For a straightened filament the MST is simply the connection of all cores along the filament. The top row of figure \ref{fig::MST_dist} shows the distribution of edge lengths for one realisation of the fiducial single-tier fragmentation (left), and the distribution from all 100 realisations (right). The distributions are strongly peaked and narrow, thus able to detect the underlying characteristic separation.

For the distribution from the single realisation: $\mathcal{M}$=0.367 pc with $\mathcal{I}$=0.041 pc, and $M$=0.355 pc with $S$=0.059 pc. The median/mean is closer to the underlying characteristic spacing, 0.4 pc, than that found by using the nearest neighbour separation. For the distribution of edge lengths from all 100 realisations: $\mathcal{M}$=0.395 pc with $\mathcal{I}$=0.129 pc, and $M$=0.396 pc with $S$=0.098 pc. Thus, the MST is able to recover the underlying spacing distribution better than the nearest neighbour separation method. This is because the nearest neighbour method has a preference for shorter separations, and some separations may be counted twice if two cores are each others nearest neighbour. This is not the case for the MST method.

The bottom row of figure \ref{fig::MST_dist} shows the distribution of edge lengths from one realisation of the fiducial two-tier fragmentation (left), and the distribution from all 100 realisations (right). The distribution from the single realisations is clearly bimodal, unlike that recovered using the nearest neighbour separation method. The peaks are located at $\sim$ 0.08 pc and $\sim$ 0.33 pc, close to the underlying characteristic lengths, 0.08 pc and 0.4 pc. The distribution of edge lengths from all 100 realisations is also clearly bimodal, with peaks $\sim$ 0.12 pc and $\sim$ 0.4 pc, but unlike the results from the nearest neighbour method, the two peaks have comparable heights, similar to the underlying PDF. 

The minimum spanning tree method is similar to the nearest neighbour method but does not have its bias to smaller separations. This leads the minimum spanning tree method to better estimate the underlying characteristic length-scale and to more robustly detect the presence of multiple length-scales.

\subsection{N$^{th}$ nearest neighbour separation}\label{SSEC:RES:NSS}%

\begin{figure*}
\includegraphics[width=0.48\linewidth]{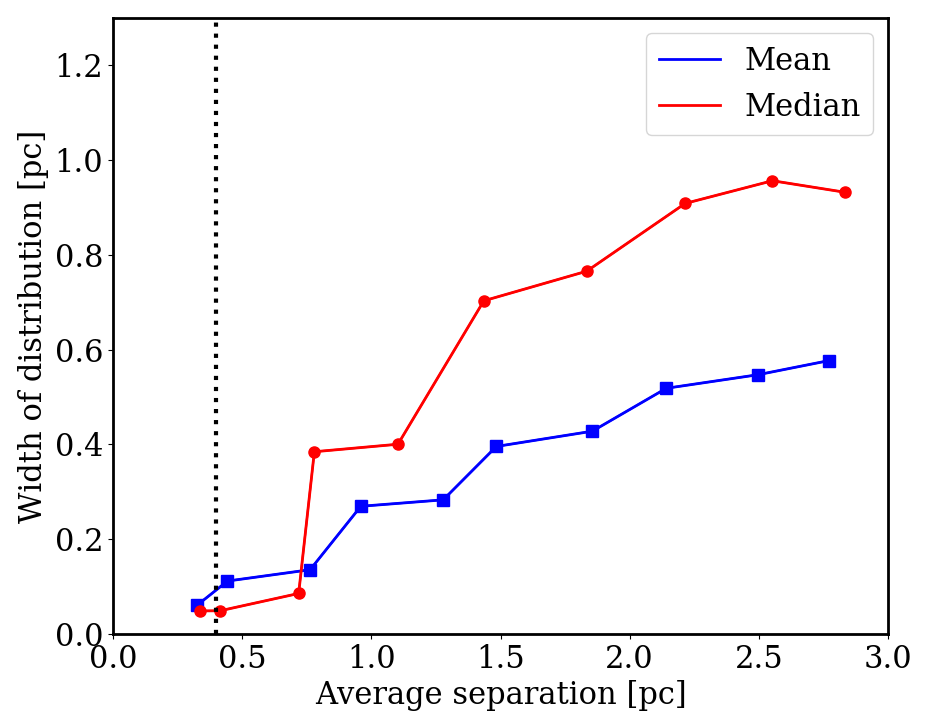}
\includegraphics[width=0.48\linewidth]{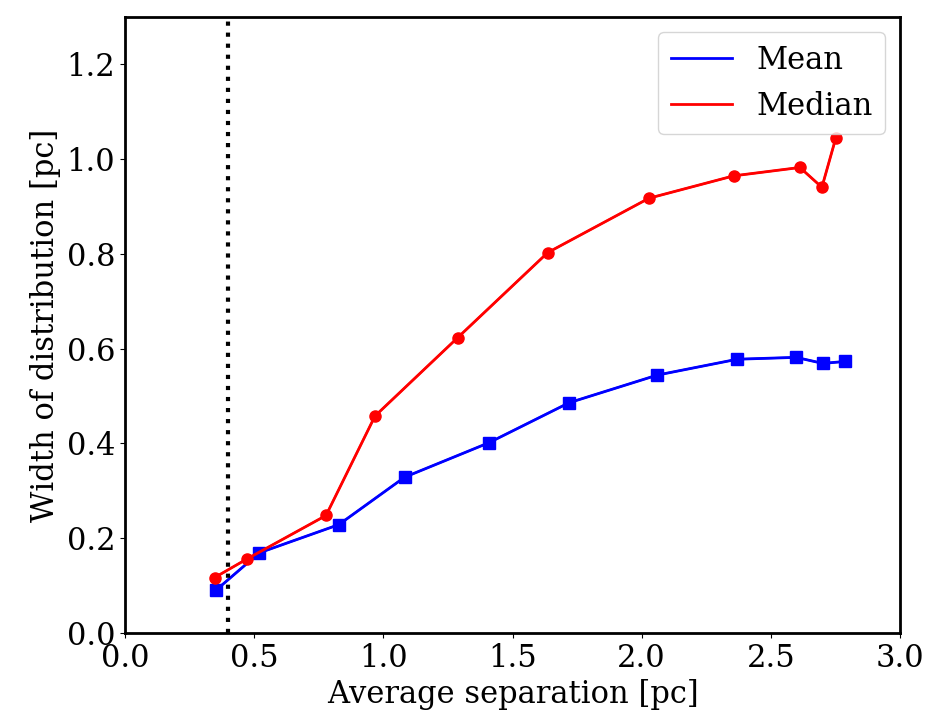}
\includegraphics[width=0.48\linewidth]{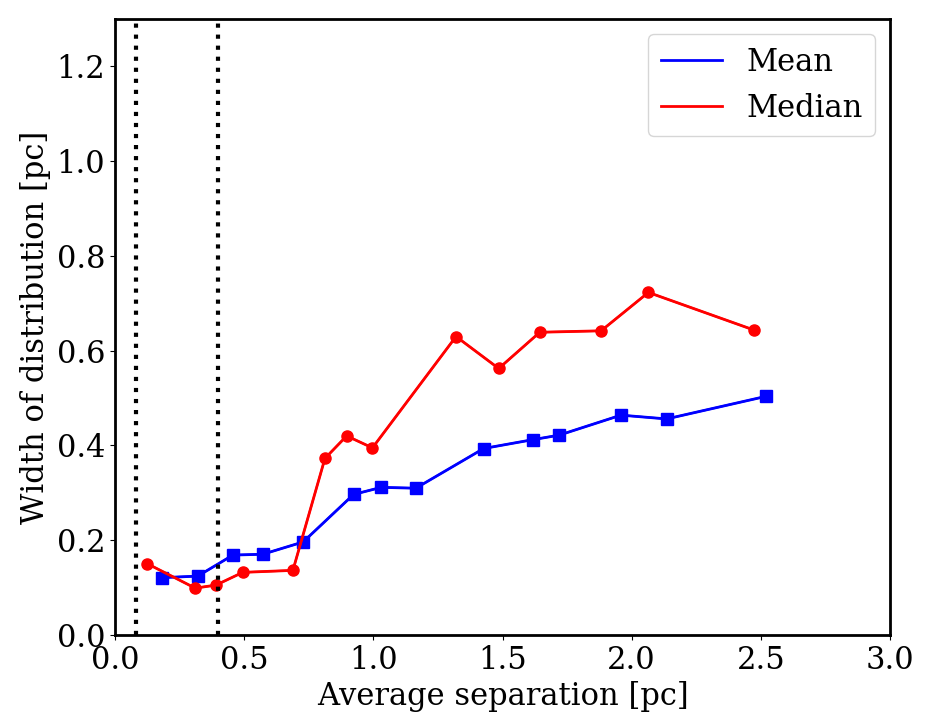}
\includegraphics[width=0.48\linewidth]{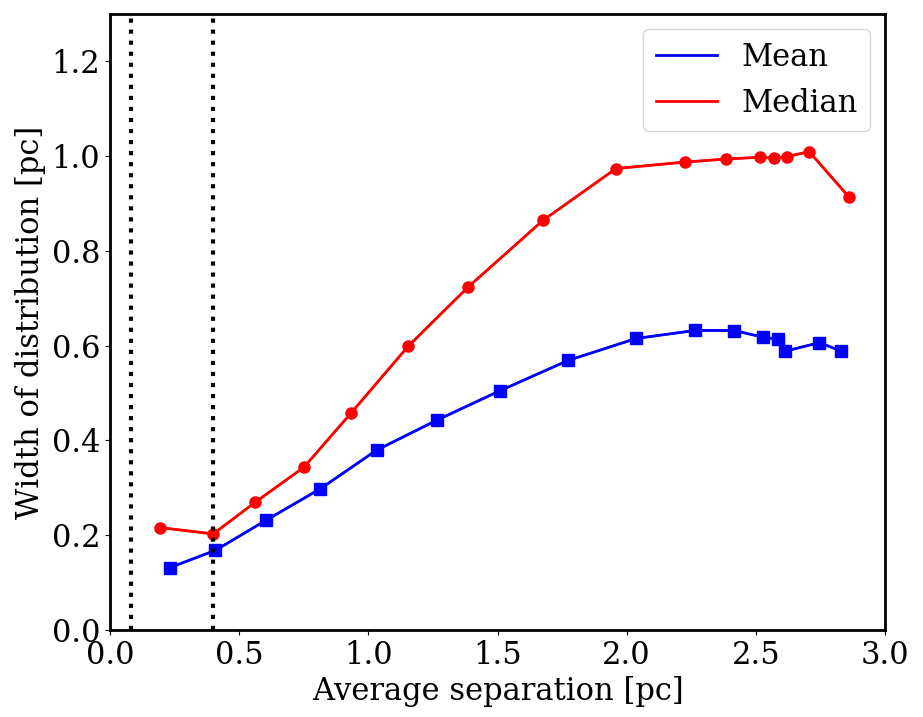}
\includegraphics[width=0.48\linewidth]{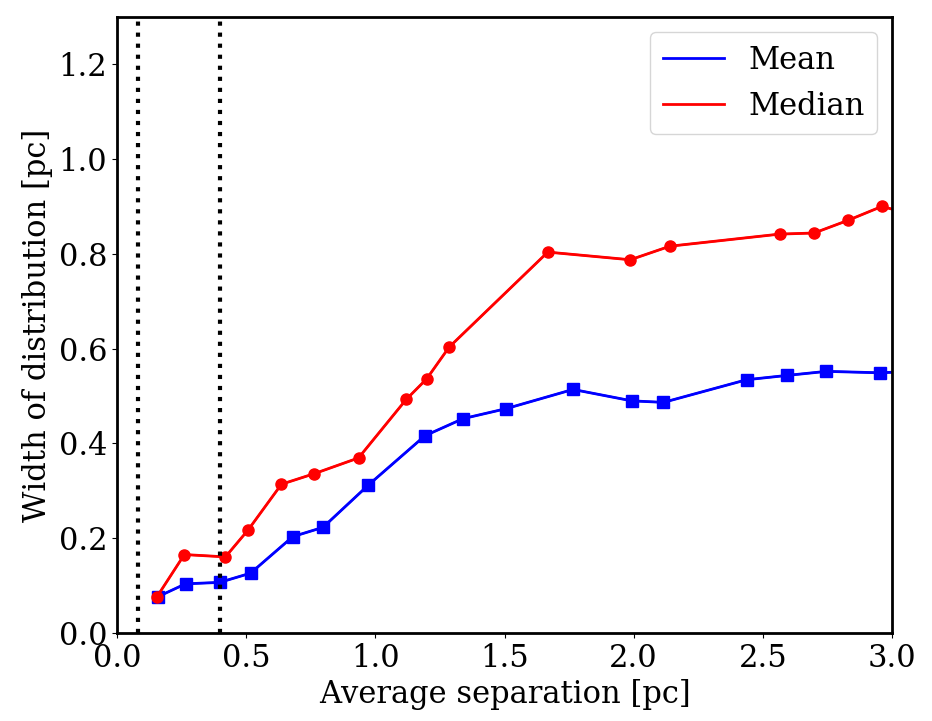}
\includegraphics[width=0.48\linewidth]{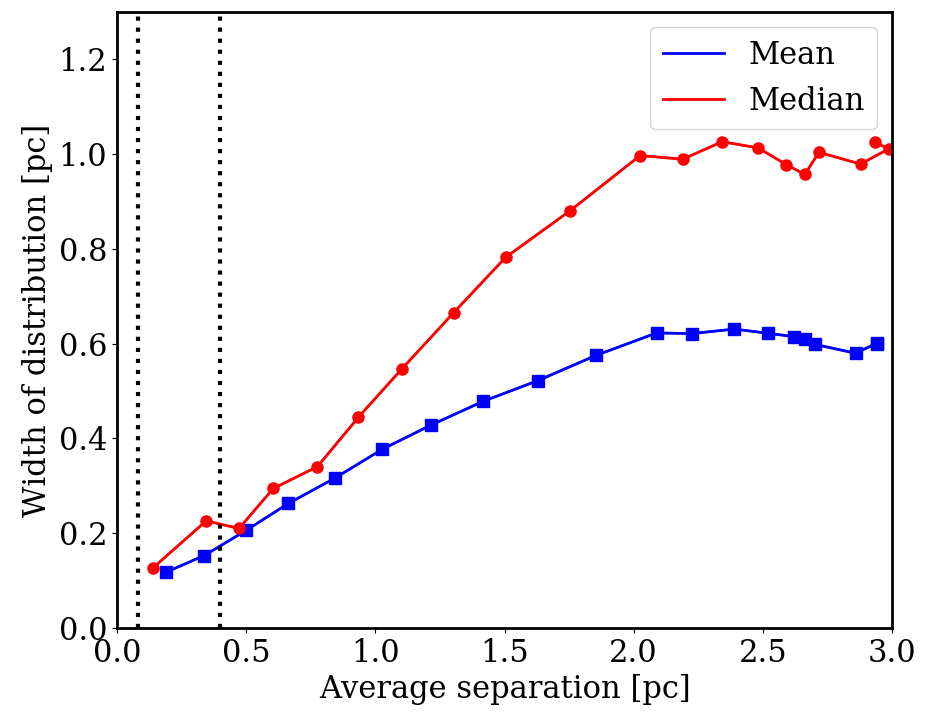}
\caption{(Left column) The results from the N$^{th}$ nearest neighbour method from a single realisation. (Right column) The results when considering the separations from all 100 realisations of the same parameters as the left column. (Top row) The fiducial single-tier fragmentation case. (Middle row) The fiducial two-tier fragmentation case. (Bottom row) The two-tier fragmentation case where the scatter, $\sigma$ in the underlying PDF of spacings is reduced to 0.04 pc. The black dotted line shows the characteristic spacings used to construct the filament, 0.4 pc for the large-scale spacing and 0.08 pc for the small-scale spacing in the two-tier case.}
\label{fig::Nth_dist}
\end{figure*}

The N$^{th}$ nearest neighbour separation technique returns an $M$ by $M$ array for $M$ cores where the N$^{th}$ column is the separation between each core and its N$^{th}$ neighbour. Each column can be reduced to an average and width; either the mean and standard deviation (blue lines) or the median and interquartile range (red lines). The top row of figure \ref{fig::Nth_dist} shows the average plotted against the width for one realisation of the fiducial single-tier fragmentation (left), and the results from all 100 realisations (right). To obtain the distribution from all 100 realisations the distributions of separations of each N$^{th}$ neighbour were combined and then averaged (rather than the distribution of each realisation's average N$^{th}$ neighbour separation being averaged). 

As there is a single characteristic wavelength for fragmentation, there should exist a single `minimum' which is the nearest neighbour. Both the median and mean show that the 1$^{st}$ neighbour separation is the neighbour with the smallest width distribution. Increasing neighbour number have wider distribution of separations. However, the median shows suppressed interquartile ranges until the 4$^{th}$ nearest neighbour which one may, incorrectly, interpret as a broad range of characteristic fragmentation length-scales. In other realisations, when using the median, there exist local minima which would be identified as characteristic spacings. Thus, the median-interquartile range can be misleading when using the N$^{th}$ nearest neighbour method.

The middle row of figure \ref{fig::Nth_dist} shows the same plots for the fiducial two-tier fragmentation realisations. Here, one would expect to see minima at both, the short (0.08 pc) and long (0.4 pc) characteristic spacings. These minima are not present. Instead, the two-tier fragmentation plots look very similar to the single-tier fragmentation plots. The bottom row of figure \ref{fig::Nth_dist} shows the same plots for the two-tier fragmentation realisations where the scatter of the input distribution, $\sigma$, is 2.5 times lower (the bottom-right plot of figure \ref{fig::synfil} is a single realisation with these parameters). Here there are tentative signs of a shallow local minimum at 0.4-0.5 pc in both the single realisation (left) and all 100 realisations (right) plots when considering the median. However as noted above, the median-interquartile range can be misleading and produce spurious minima which do not correspond to a real characteristic length-scale. The mean and standard deviation shows no strong indication of two-tier fragmentation. 

Despite containing information on numerous scales the N$^{th}$ nearest neighbour method appears insensitive to multi-scale fragmentation. This can be understood by the fact that the distributions of separations to each N$^{th}$ neighbour are reduced to their average and width. As seen above in section \ref{SSEC:RES:NNS}, the average and width of a bimodal distribution, like the separation distributions obtained when multi-scale fragmentation exists, are poor descriptors of the distribution. 

We conclude that the N$^{th}$ nearest neighbour method is rather insensitive to multi-scale fragmentation. Moreover, when compared to the nearest neighbour method, no new information about the presence of a single characteristic fragmentation length-scale is gleaned from the use of the N$^{th}$ nearest neighbour method. For these reasons we do not favour its use.

\subsection{Two-point correlation function}\label{SSEC:RES:2P}%

\begin{figure*}
\includegraphics[width=0.48\linewidth]{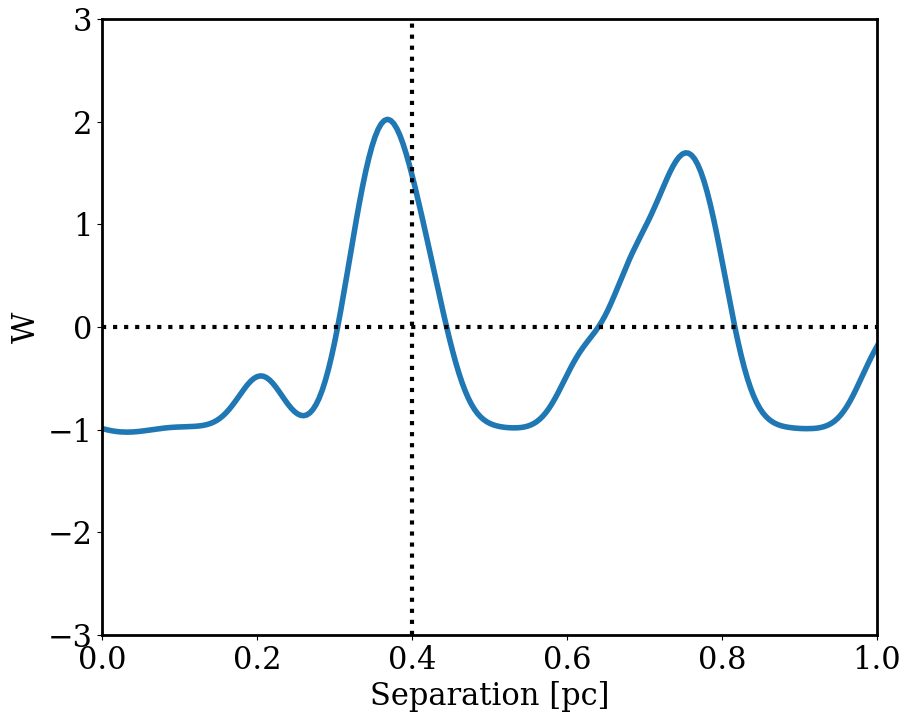}
\includegraphics[width=0.48\linewidth]{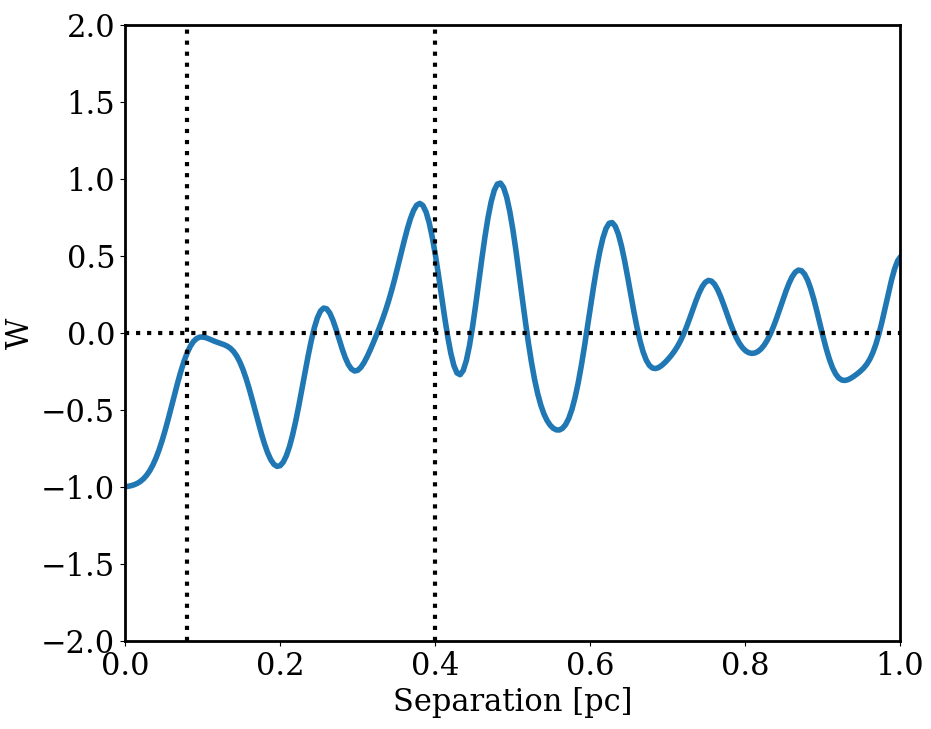}
\caption{The two-point correlation function in the range $0\leq x \leq 1$ from a single realisation of (left) the fiducial single-tier fragmentation case and (right) the fiducial two-tier fragmentation case. The vertical black dotted lines show the characteristic spacings used to construct the filament, 0.4 pc for the large-scale spacing and 0.08 pc for the small-scale spacing in the two-tier case. The horizontal black dotted line shows $y=0$ to help the reader distinguish between features which are significant and those which are not.}
\label{fig::2Point}
\end{figure*} 

\begin{figure*}
\includegraphics[width=0.47\linewidth]{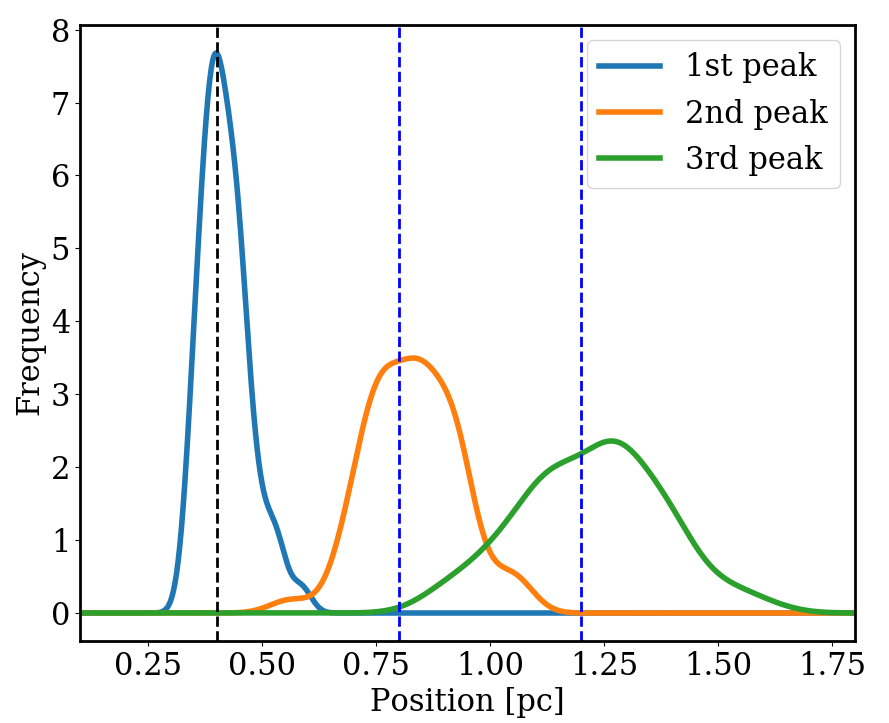}
\includegraphics[width=0.47\linewidth]{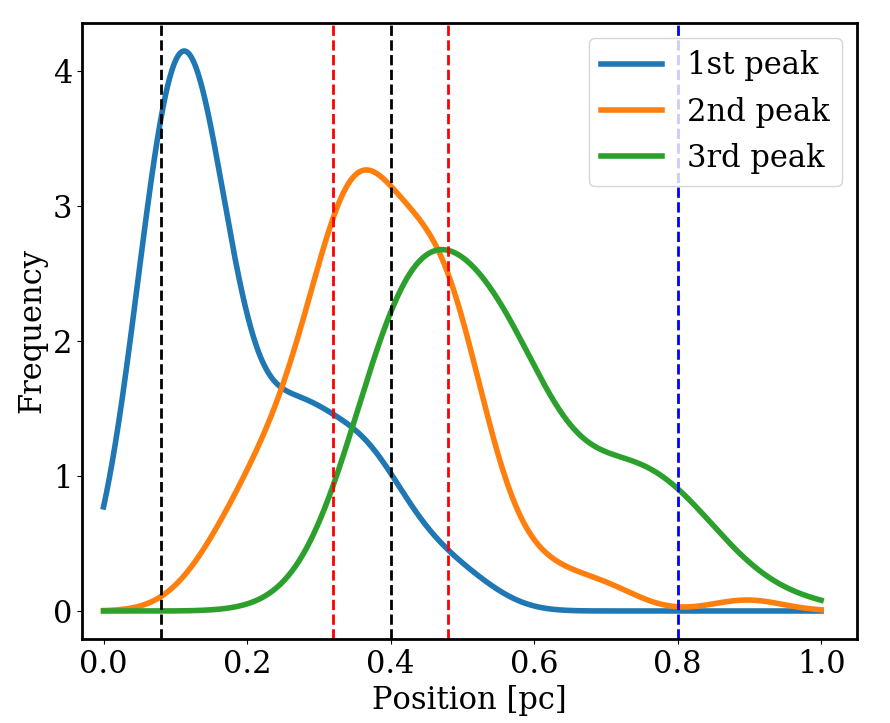}
\caption{KDEs showing the distribution of the first three significant peak locations of the two-point correlation functions resulting from all 100 realisations of the fiducial single-tier (left) and two-tier (right) fragmentation parameters. The vertical black dashed lines show the underlying characteristic fragmentation length-scales and the vertical blue dashed line show the harmonics of the characteristic length-scale. For the right plot the vertical red dashed lines show the superpositions of the two characteristic length-scales. }
\label{fig::peakkde}
\end{figure*} 

\begin{figure}
\includegraphics[width=0.95\linewidth]{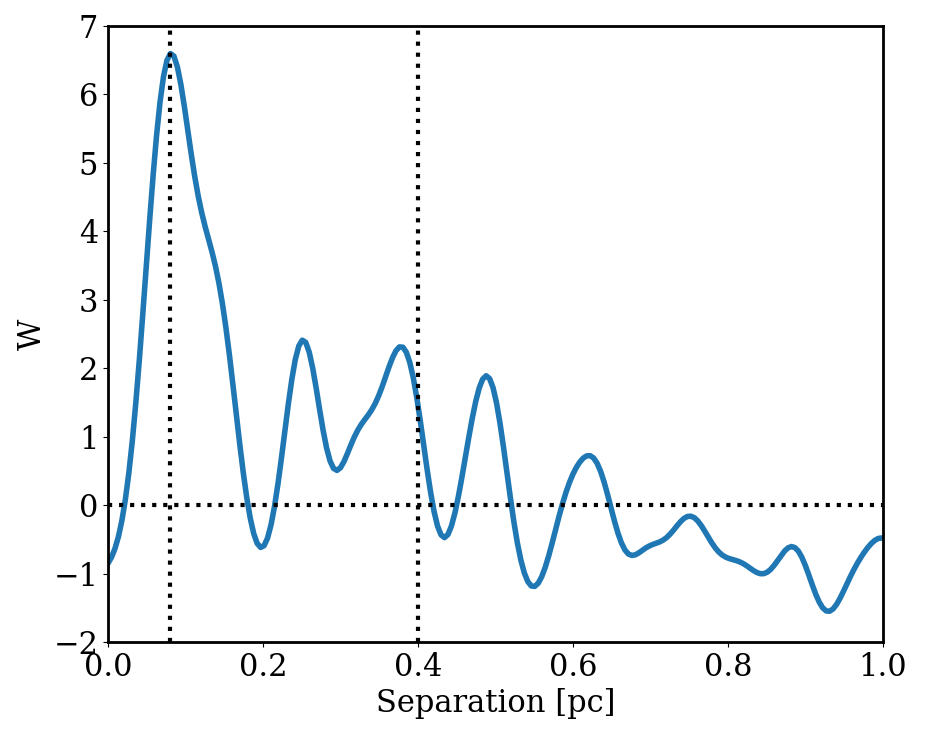}
\caption{The two-point correlation function in the range $0\leq x \leq 1$ from the same realisation of the fiducial two-tier fragmentation case shown in figure \ref{fig::2Point} when the randomly placing cores are placed anywhere in the map. The dotted lines are the same as in figure \ref{fig::2Point}}
\label{fig::2Point_wrong}
\end{figure}

The two-point correlation function is a continuous function made from a set of discrete points, i.e the separations between every pair of data points. A histogram or a Kernel Density Estimator (KDE) is used for this purpose. A KDE converts a set of discrete points into a continuous distribution by convolving each data-point with a kernel, in this case a Gaussian with a given width. In this work we use a KDE to produce the two-point correlation function due to its smoothness. The width of the kernel is determined using Scott's method \citep{Sco92}. The $RR(r)$ and $DR(r)$ are produced from 100,000 instances of randomly placed cores along the filament. Due to the large number of data points needing to be converted into a KDE, an approximate KDE using a tree structure is used. The evaluation of these approximate KDEs is between 50 to 100 times faster than the exact evaluation while errors are less than $0.1\%$. Details can be found in appendix \ref{APP:FRAG:AKDE}. 

The left panel of figure \ref{fig::2Point} shows the two-point correlation function of a single realisation with the fiducial single-tier fragmentation parameters. The determined width of the kernel is $\sim \, 0.03 \, \rm pc$. The first significant peak (i.e. with a value above 0) is located at $\sim 0.35 \rm pc$, close to the characteristic spacing of 0.4 pc. The signal is sufficiently strong that the higher order resonances are seen at $\sim$ 0.8, 1.2, 1.6 etc.. The left panel of figure \ref{fig::peakkde} shows the position of the first 3 peaks which are statistically significant ($W>0$) from the two-point correlation functions resulting from the 100 realisations using the fiducial single-tier fragmentation parameters. The first peak has a narrow dispersion around the characteristic fragmentation length-scale at 0.4 pc. The next two peaks lie around the first and second harmonics at 0.8 pc and 1.2 pc. However, there exist significant scatter, showing that for numerous realisations the signal from the harmonics may not be strong and spurious peaks unrelated to the characteristic fragmentation scale may appear.  

The right panel of figure \ref{fig::2Point} shows the results for a single realisation using the fiducial two-tier fragmentation parameters. Here, the two-point correlation function is far more complicated than the single tier case, showing multiple peaks which are not perfectly periodic. A peak exists at $x_1 \sim 0.1 \, \rm pc$ where $W \sim 0$, but not quite, and it corresponds to the small-scale spacing at 0.08 pc. There also exists a significant peak at $x_3 \sim 0.35 \, \rm pc$ which corresponds to the large-scale spacing at 0.4 pc. Between these two peaks is a smaller, but still significant, peak at $\sim 0.25 \, \rm pc$. This is a superposition between the first and third peak, $x_3 - x_1$. Another such superposition occurs at $\sim 0.45 \, \rm pc$, at $x_3 + x_1$. Without knowledge of the true characteristic spacings it is non-trivial to determine which of these peaks correspond to the underlying fragmentation length-scales. 

As seen in the two panels of figure \ref{fig::2Point}, the first few peaks are the only ones in the two-point correlation function which contain information on the characteristic filament fragmentation length-scales, later peaks are resonances and superpositions of the characteristic length-scales. 

The right panel of figure \ref{fig::peakkde} shows the position of the first 3 peaks which are statistically significant ($W>0$) from the two-point correlation functions resulting from the 100 realisations using the fiducial two-tier fragmentation parameters. The first significant peak aligns well with the small-scale spacing at 0.08 pc, and in those cases where this peak is not significant the first peak is found at the separation corresponding to the first superposition at 0.32 pc. The second peak in the two-point correlation function is well aligned with either the large-scale characteristic spacing at 0.4 pc, or one of the two superpositions at 0.32 and 0.48 pc. The third significant peak in the two-point correlation function is associated with the large-scale characteristic spacing, the upper superposition at 0.48 pc, or the first resonance of the large-scale characteristic spacing at 0.8 pc. Thus, the positions of the peaks of the two-point correlation function contain physically meaningful information but can be difficult to interpret. 

When there exists a single characteristic fragmentation length-scale, the first significant peak corresponds to it. But when there are two characteristic length-scales ($x_1,x_2$) then one must consider the superposition between them, $x_2 - x_1$, meaning the first 3 peaks should be examined. For three characteristic length-scales ($x_1,x_2,x_3$) there can be up to 5 superpositions which may be significant, $x_2 - x_1, x_2 + x_1, x_3 - x_2 - x_1, x_3 - x_2, x_3 - x_1$, meaning that the first 8 peaks must be considered. An observer has no prior knowledge of the number of characteristic fragmentation length-scales and so must instead deduce the minimum number of such length-scales with which one can explain the peaks in the two-point correlation function resulting from their data. The results of this deduction should then be compared to the model selection described in section \ref{SSEC:Model}.

As discussed in section \ref{SSEC:2point}, the evaluation of the two-point correlation function requires a comparison with randomly placed cores. When producing the randomly placed cores, one may place them randomly along the filament (as done above) or randomly in the observed 2D map \citep[as done by ][]{Kai17}. Therefore, we perform a test by re-evaluating the two-point correlation function of the same two-tier fragmentation realisation as shown in the right panel but randomly placing cores anywhere in the map (see figure \ref{fig::2Point_wrong}). This leads to an enhancement at small-scale separations. Large-scale separations, $\sim 1$ pc, show a corresponding deficiency. Using this method therefore accentuates the statistical significance of small-scale separations in the sample at the price of larger-scale ones, and should thus be avoided. Moreover, as the hypothesis one wishes to test is if there exists a characteristic \textit{filament} fragmentation length-scale one should strictly place the cores only along the filament.

The two-point correlation function appears to be a powerful tool with its ability to robustly detect single and multiple characteristic length-scales. However, its sensitivity causes complexity in its appearance and care must be taken in its interpretation. It should therefore be used in conjunction with the simpler minimum spanning tree and nearest neighbour methods. One must also take care with the placement of randomly placed cores to ensure one is testing the correct fragmentation scenario.

\subsection{Fourier power spectrum}%

\begin{figure*}
\includegraphics[width=0.45\linewidth]{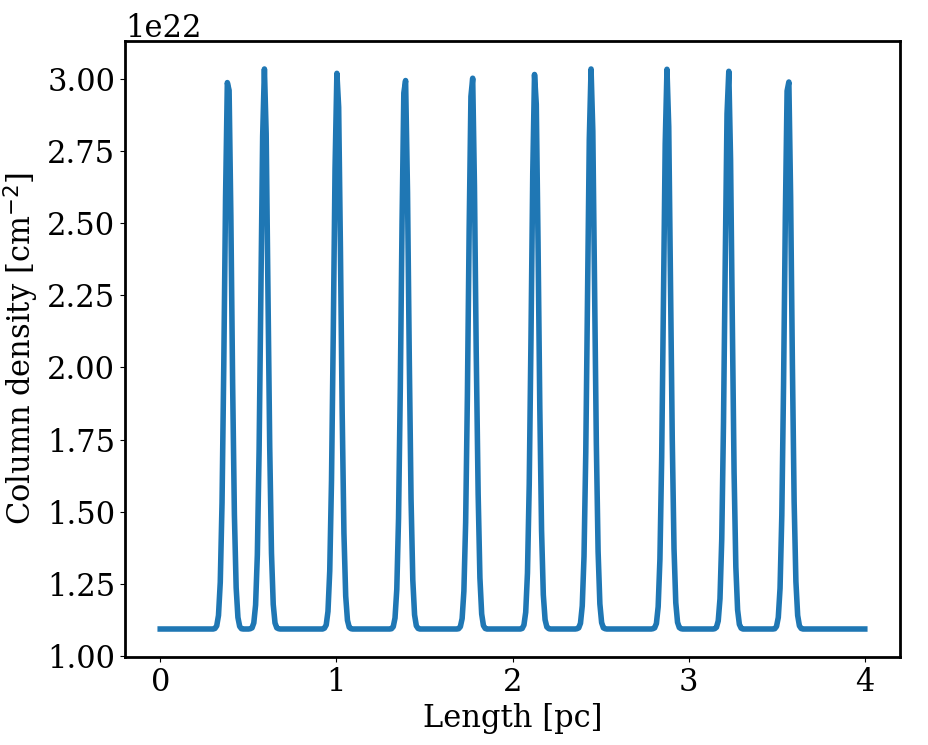}
\includegraphics[width=0.45\linewidth]{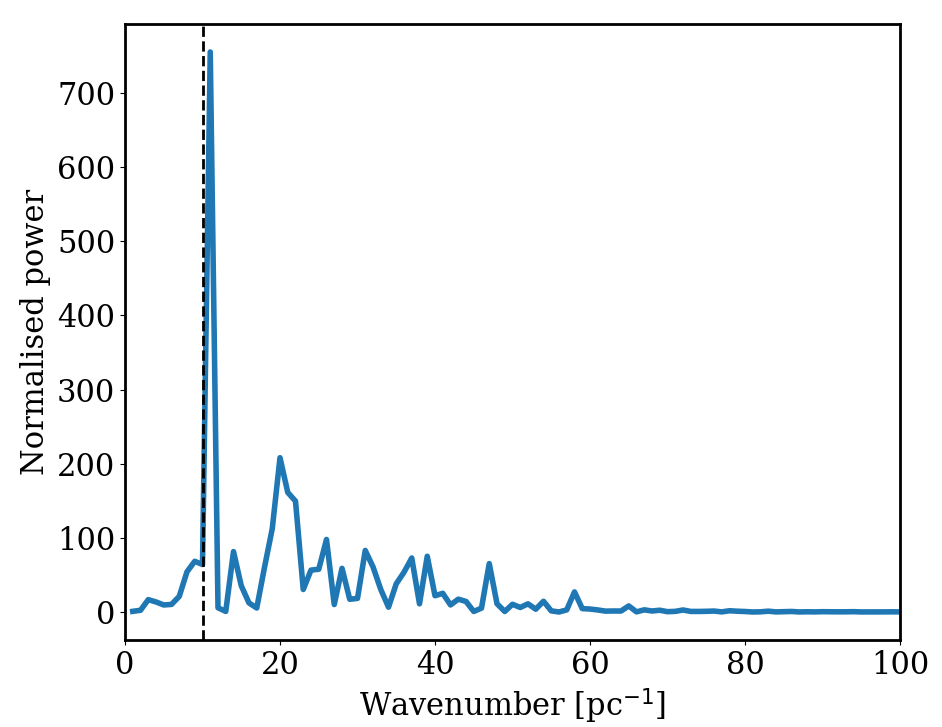}
\includegraphics[width=0.45\linewidth]{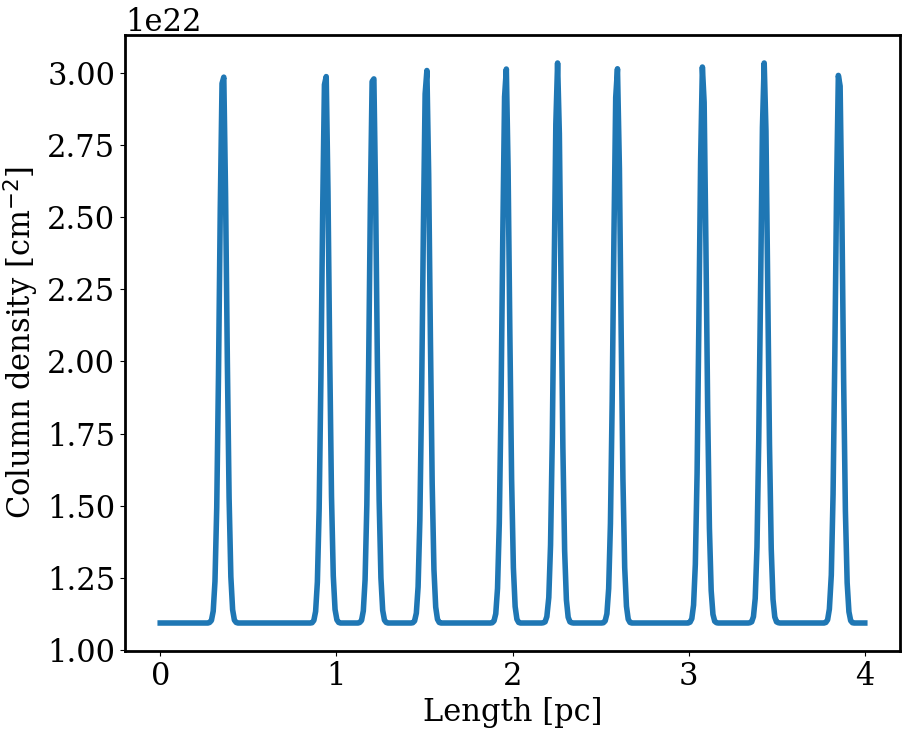}
\includegraphics[width=0.45\linewidth]{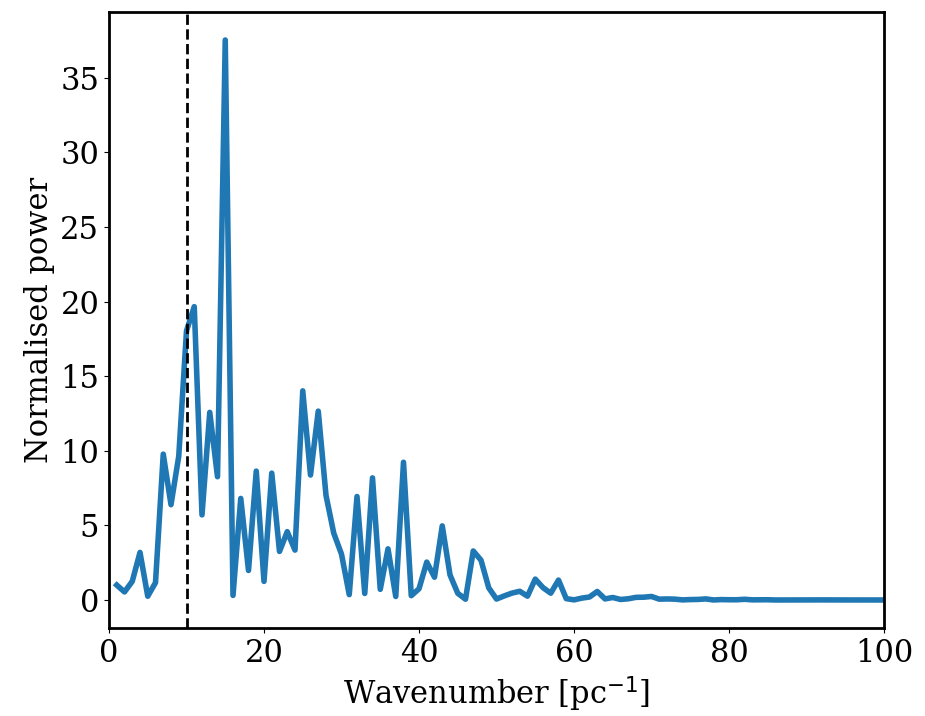}
\includegraphics[width=0.45\linewidth]{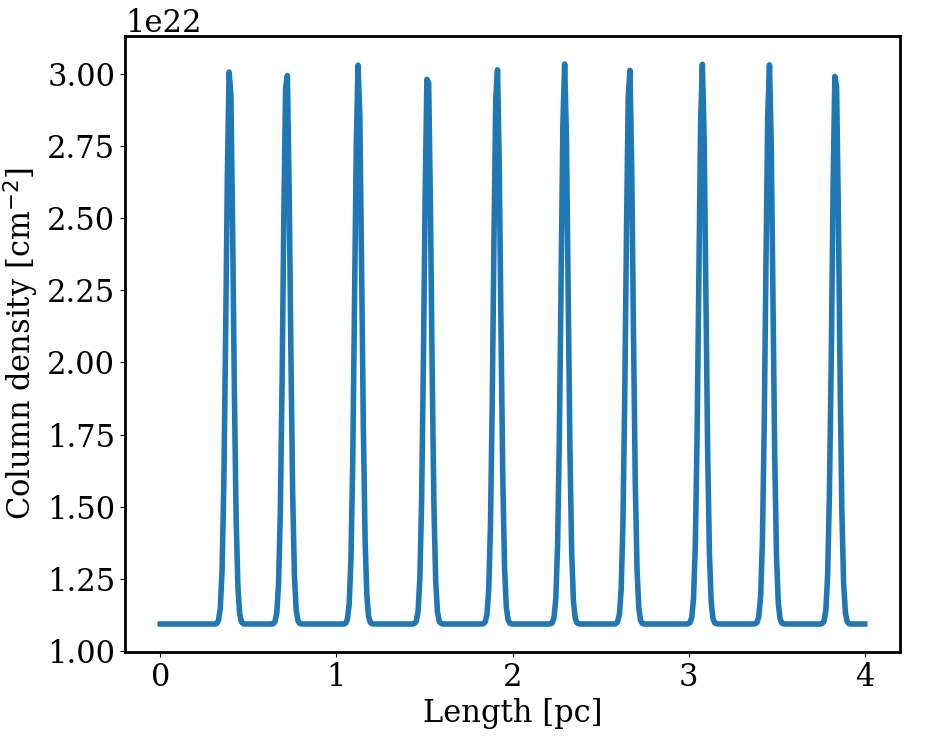}
\includegraphics[width=0.45\linewidth]{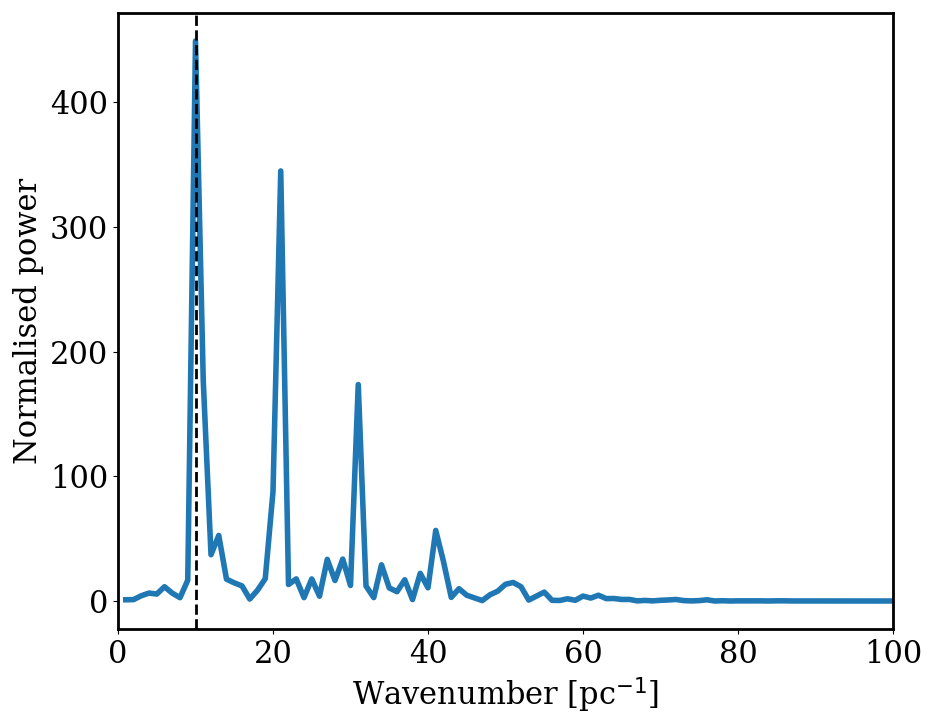}
\caption{(Left column) The column density profile along the spine of synthetic filaments constructed using single-tier fragmentation parameters. The top and middle rows are two realisations with the fiducial set of parameters and the bottom with the scatter $\sigma = 0.04$ pc. (Right column) The respective power spectrum, normalised such that $P(k=1)=1.0$. The vertical black dashed line at $k=10$ corresponds to the characteristic fragmentation length-scale at 0.4 pc.}
\label{fig::FTs}
\end{figure*} 

\begin{figure*}
\includegraphics[width=0.45\linewidth]{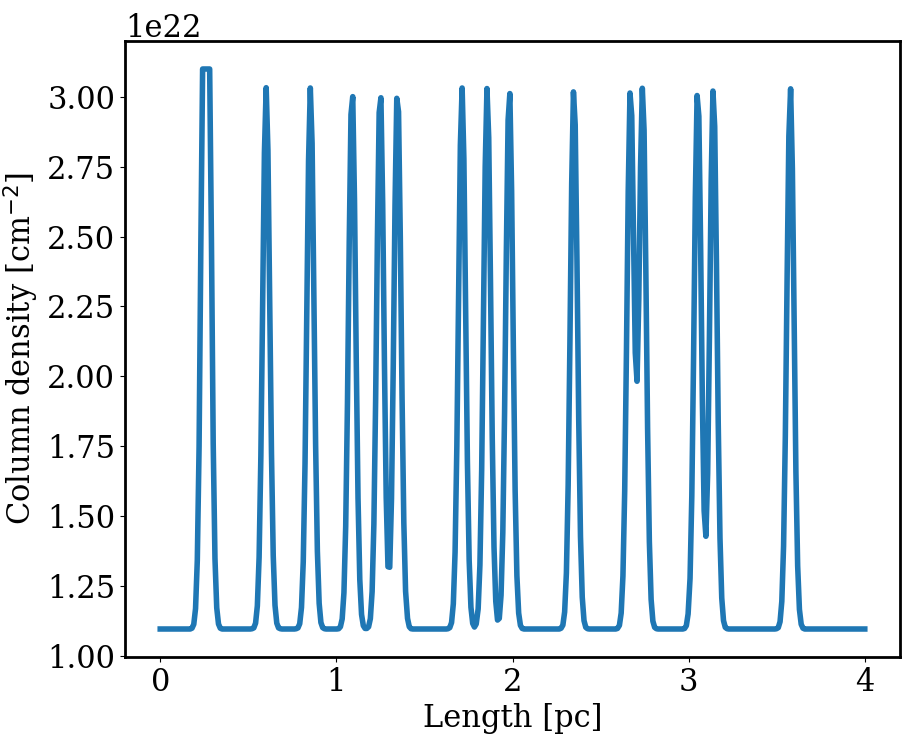}
\includegraphics[width=0.45\linewidth]{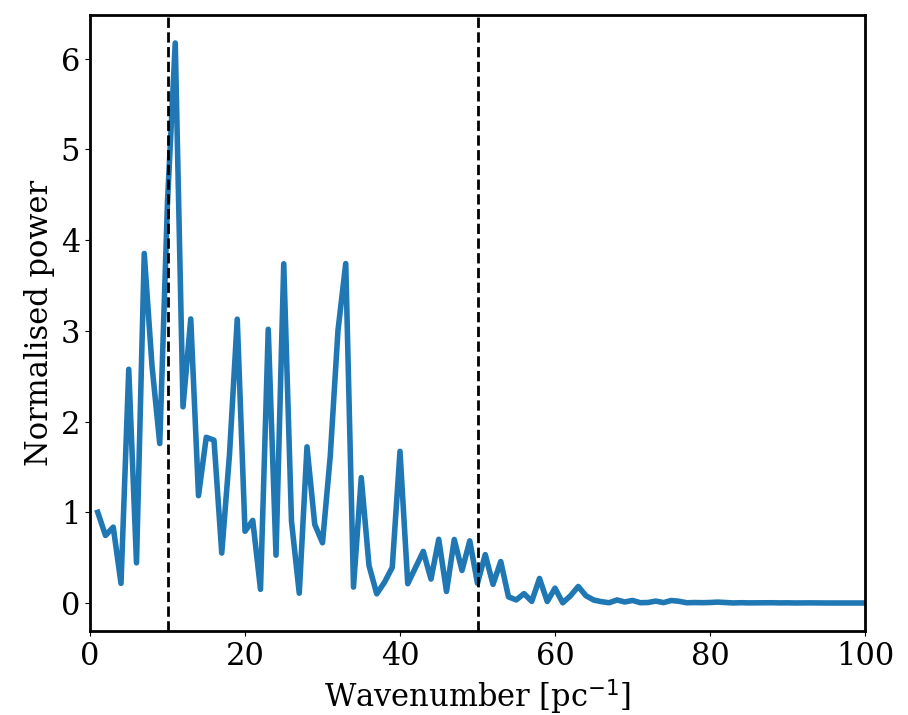}
\caption{As in figure \ref{fig::FTs} but for a synthetic filament constructed using the fiducial two-tier fragmentation parameters.}
\label{fig::FTs-double}
\end{figure*} 

The Fourier transform does not require the core positions, but rather just the spine of the filament and the column density map. The top and middle panels of figure \ref{fig::FTs} show the column density profile along the filament spine for two realisations using the fiducial single-tier fragmentation parameters, and the corresponding normalised power spectrum of the Fourier transform. The power spectrum has been normalised such that $P(k=1) = 1.0$ where $k=1$ corresponds to the length of the filament, 4 pc. The top realisation is the one shown throughout this paper (see the leftmost panel of figure \ref{fig::synfil}). The power spectrum in the top row shows a large peak at $k \sim 10$ which corresponds to the characteristic spacing of 0.4 pc. However, it also shows numerous other peaks, which if one does not know the number of characteristic length-scales, could be interpreted as such. The middle row of figure \ref{fig::FTs} shows a second realisation using the same fiducial single-tier fragmentation parameters, here the largest peak occurs at $k \sim 15$, corresponding to a length-scale of $\sim 0.25$ pc. The rather small scatter ($\sigma = 0.1$ pc for the fiducial set of parameters) in the underlying characteristic spacing is sufficient to produce considerable noise in the power spectrum. Reducing the scatter to 0.04 pc, as seen in the bottom row of figure \ref{fig::FTs}, produces a power spectrum which is strongly peaked only at the characteristic length-scale and its resonances. 

Figure \ref{fig::FTs-double} shows the column density profile along the filament spine for a single realisation using the fiducial two-tier fragmentation parameters (left), and the normalised power spectrum (right). While the largest peak is at $k=10$, corresponding to the 0.4 pc large-scale characteristic spacing, it lies within a forest of peaks extending to $k \sim 40$. There is no evidence of a peak at $k = 50$, corresponding to the small-scale fragmentation spacing. Furthermore, the power spectrum appears similar to those seen in figure \ref{fig::FTs} where the fragmentation being modelled is only single-tier. 

We conclude that the Fourier transform can robustly detect characteristic fragmentation length-scales only when there is little scatter; a situation unlikely to occur in real observations.

\begin{table*}
\centering
\begin{tabular}{@{}*4l@{}}
\hline\hline
Method                            & Positives                                       & Negatives                                               & Recommended      \\
                                  &                                                 &                                                         & use              \\ \hline
Nearest neighbour                 & Simple.                                         & Tendency to smaller separation                          & \checkmark       \\
separation                        & Robustly detects single                         & values.                                                 &                  \\
                                  & length-scales.                                  & Typically insensitive to multiple                       &                  \\
                                  &                                                 & length-scales.                                          &                  \\ \\
Minimum spanning tree             & Simple.                                         & -                                                       & \checkmark       \\
edge length                       & No tendency to smaller                          &                                                         &                  \\
                                  & separation values.                              &                                                         &                  \\ 
                                  & Can detect single and multiple                  &                                                         &                  \\
                                  & length-scales.                                  &                                                         &                  \\ \\
N$^{th}$ nearest neighbour        & Simple.                                         & The same as for nearest neighbour.                      & \cross           \\ 
                                  &                                                 & Spurious features can appear which                      &                  \\
                                  &                                                 & are not related to real length-scales.                  &                  \\ \\
Two-point correlation             & Robustly detects single and                     & Complex appearance when there are                       & \checkmark       \\
function                          & multiple length-scales.                         & multiple length-scales                                  &                  \\
                                  & Includes an inbuilt                             &                                                         &                  \\
                                  & null-hypothesis test.                           &                                                         &                  \\ \\ 
Fourier power spectrum            & No core identification needed                   & Only robustly detects length-scales                     & \cross           \\
                                  &                                                 & when there is little scatter around                     &                  \\
                                  &                                                 & them.                                                   &                  \\ \hline \hline 
\end{tabular}
\centering
\caption{A table summarising the positives and negatives of each of the methods, and if they ought to be used. We recommend that the nearest neighbour separation, minimum spanning tree edge lengths and the two-point correlation function ought to be used in concert as they can most robustly detect characteristic fragmentation length-scales.}
\label{tab::Pos_Neg_methods}
\end{table*}

\section{Discussion}\label{SEC:DIS}%
   
\subsection{Measuring the statistical significance of the results}\label{SUBSEC:STATS}%

One must test if the results found from one of the previously discussed methods is statistically significant. Here, we show a number of techniques included in \textsc{FragMent} which are used to address this. The results are summarised in table \ref{tab::NHT_res}.

\subsubsection{Null hypothesis testing}\label{SUBSUBSEC:NHT}%

\begin{figure*}
\includegraphics[width=0.48\linewidth,trim={5.4cm 0 2cm 2cm},clip]{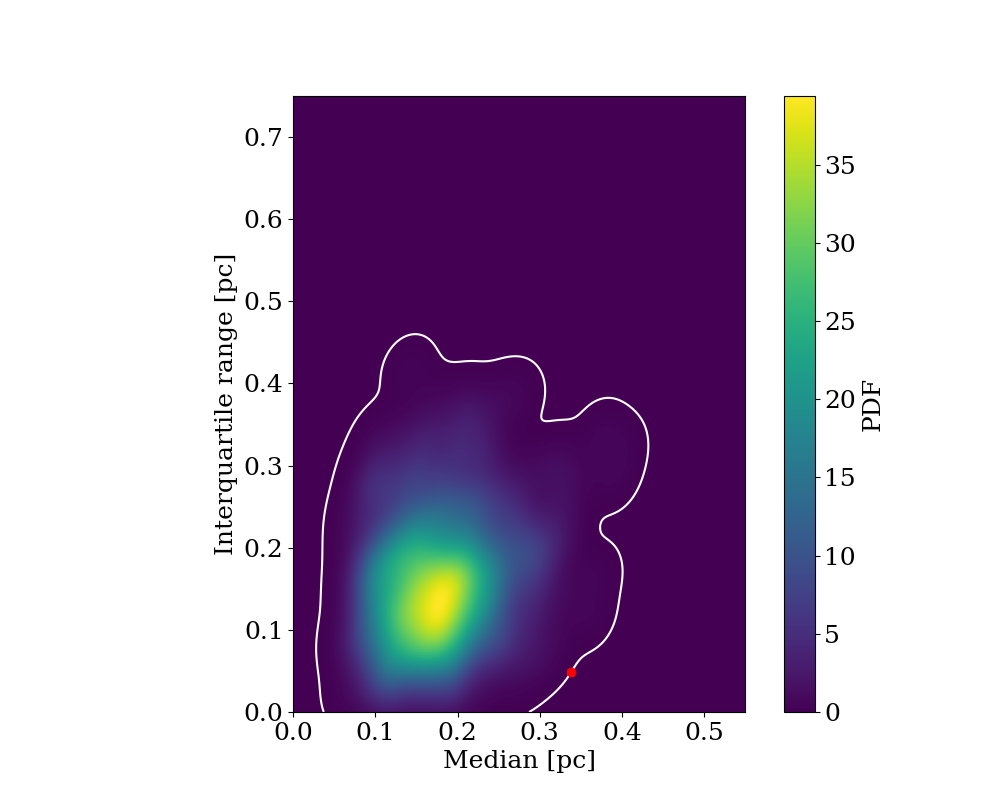}
\includegraphics[width=0.48\linewidth,trim={5.4cm 0 2cm 2cm},clip]{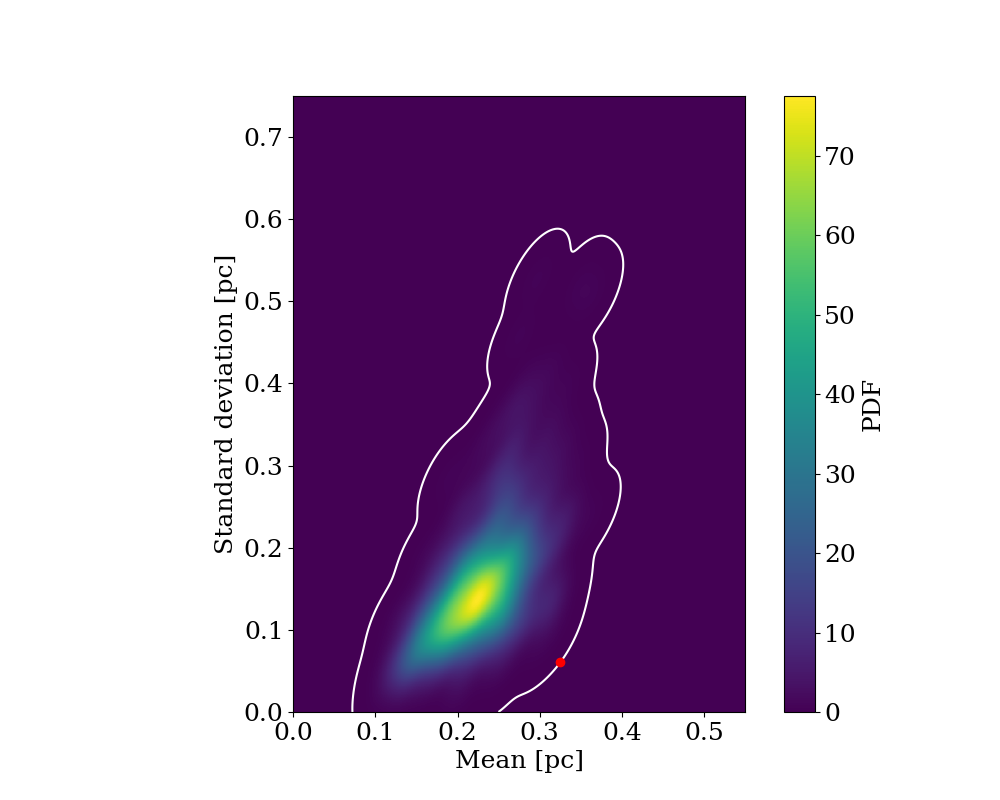}
\includegraphics[width=0.48\linewidth,trim={5.4cm 0 2cm 2cm},clip]{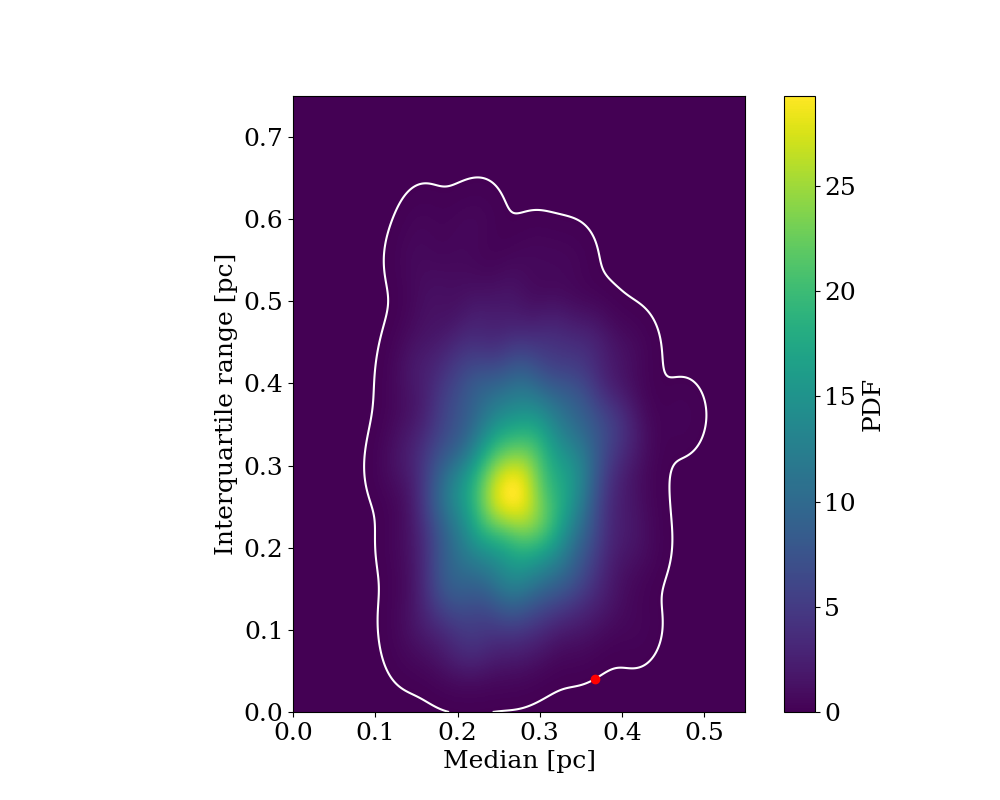}
\includegraphics[width=0.48\linewidth,trim={5.4cm 0 2cm 2cm},clip]{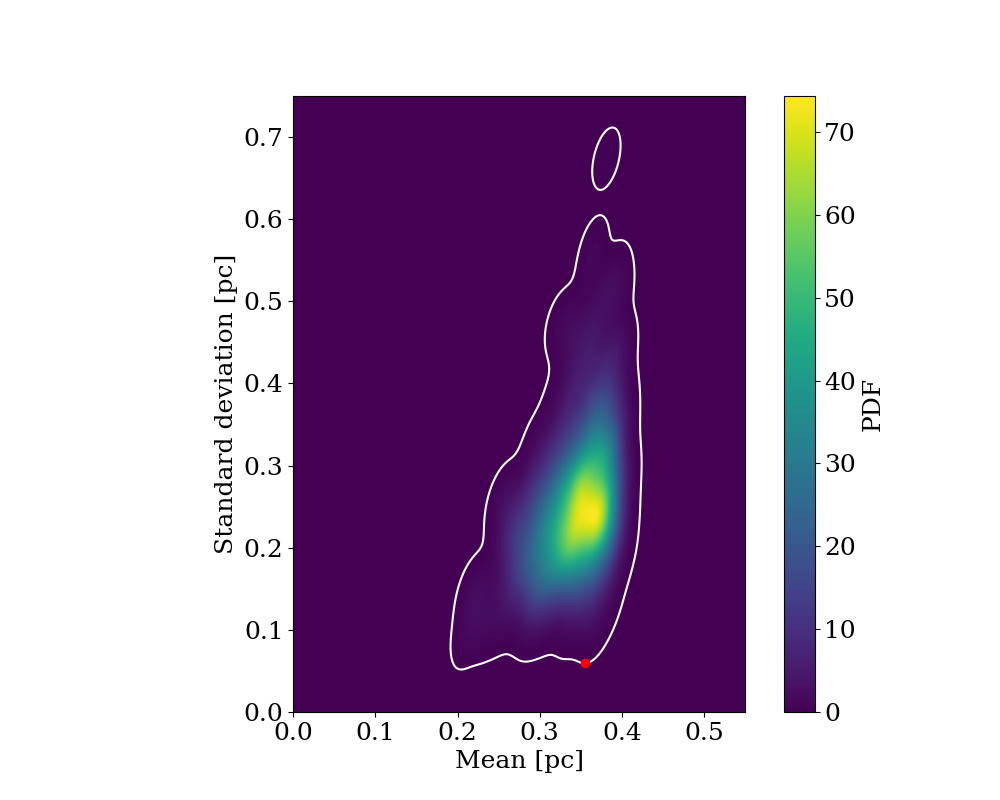}
\caption{2D PDFs showing the average and width of the separation distributions from 10,000 realisations of randomly placed cores. (Top row) The nearest neighbour method is used to find the separation distribution. (Bottom row) The minimum spanning tree method is used to find the separation distribution. (Left column) The average and width measurements are the median and interquartile range. (Right column) The average and width measurements are the mean and standard deviation. The red dot on each plot is the position of the average and width from the single-tier realisation discussed in subsections \ref{SSEC:RES:NNS} and \ref{SSEC:RES:MST}. The white contour shows all points with the same probability of occurring as the red dot.}
\label{fig::2D_PDF}
\end{figure*}

\begin{figure*}
\includegraphics[width=0.48\linewidth]{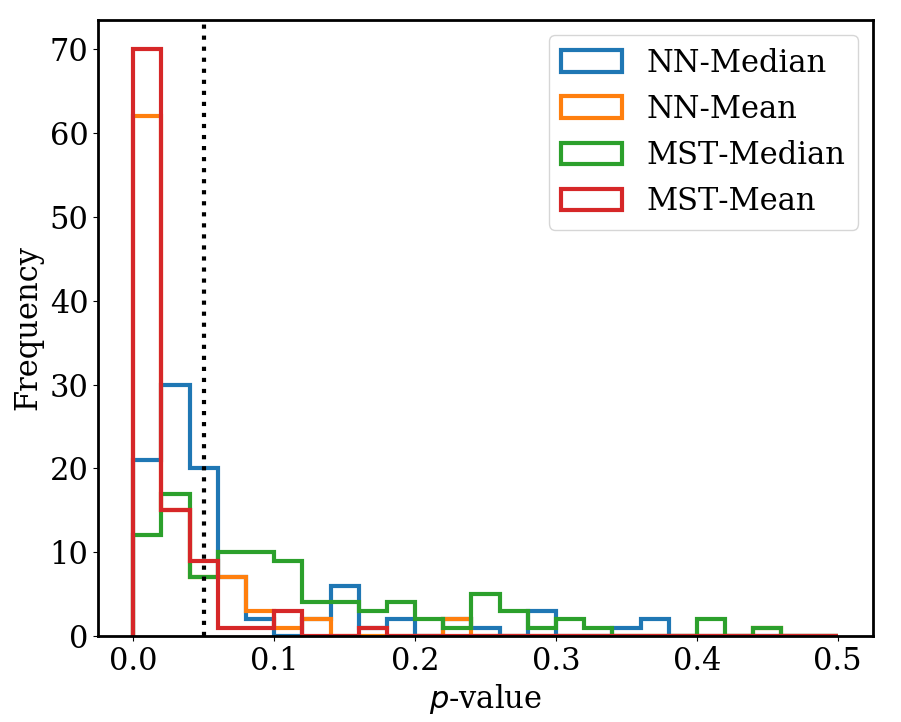}
\includegraphics[width=0.48\linewidth]{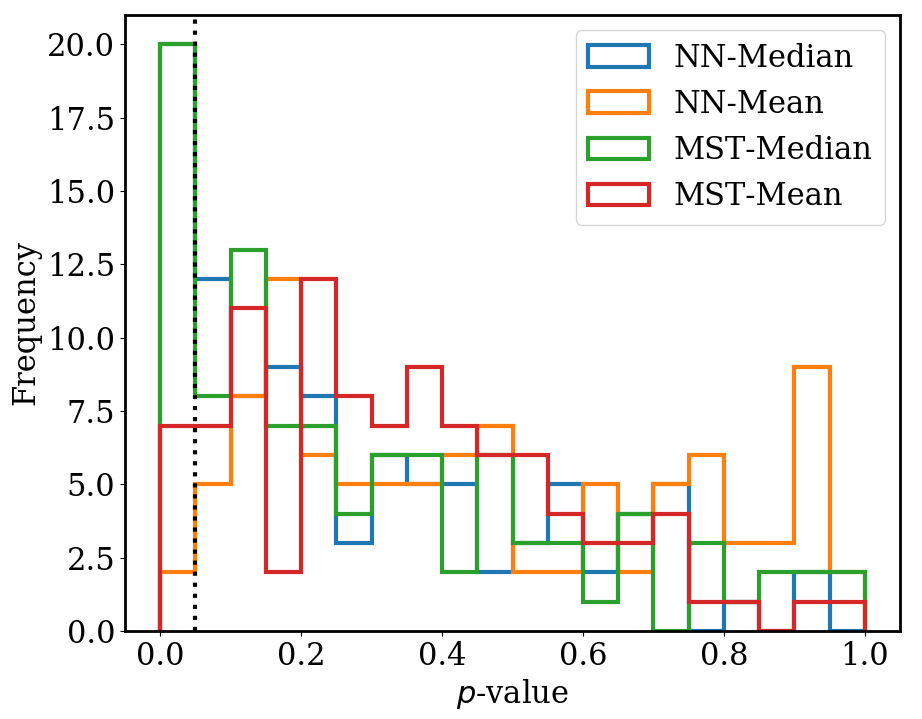}
\caption{Histograms of $p$-values from using the null hypothesis test described in section \ref{SUBSUBSEC:NHT}. (Left) The results when considering all 100 realisations with the fiducial single-tier fragmentation parameters. (Right) The results when considering all 100 realisations with the fiducial two-tier fragmentation parameters. The vertical dotted line denotes the $p$-value threshold used to determine if the null hypothesis can be rejected, 0.05. In the legend, NN stands for nearest neighbour, and MST, for minimum spanning tree.}
\label{fig::NHT_hist}
\end{figure*}

\begin{figure*}
\includegraphics[width=0.48\linewidth]{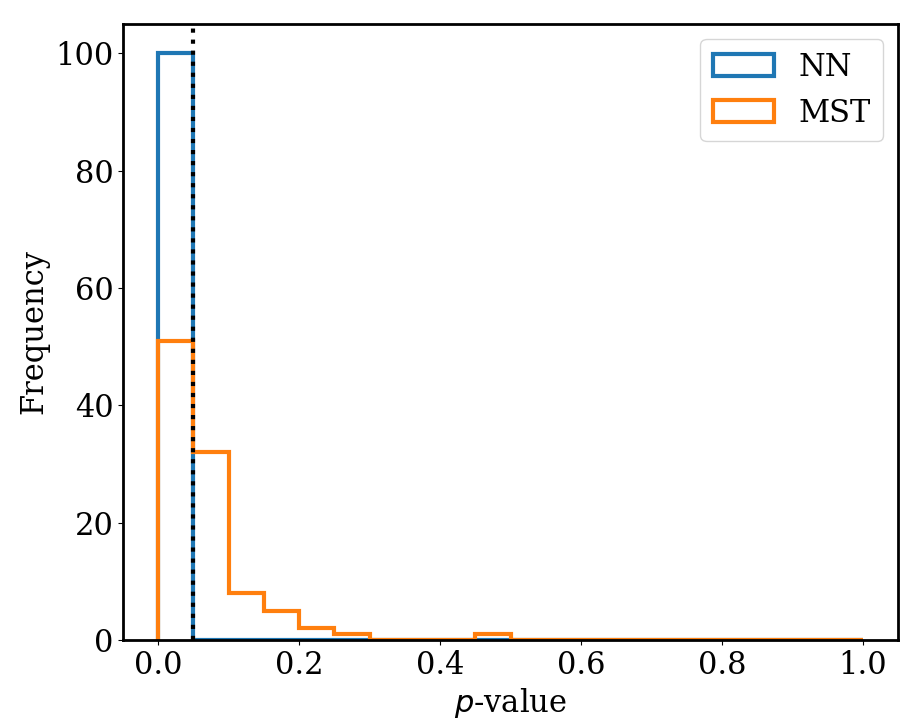}
\includegraphics[width=0.48\linewidth]{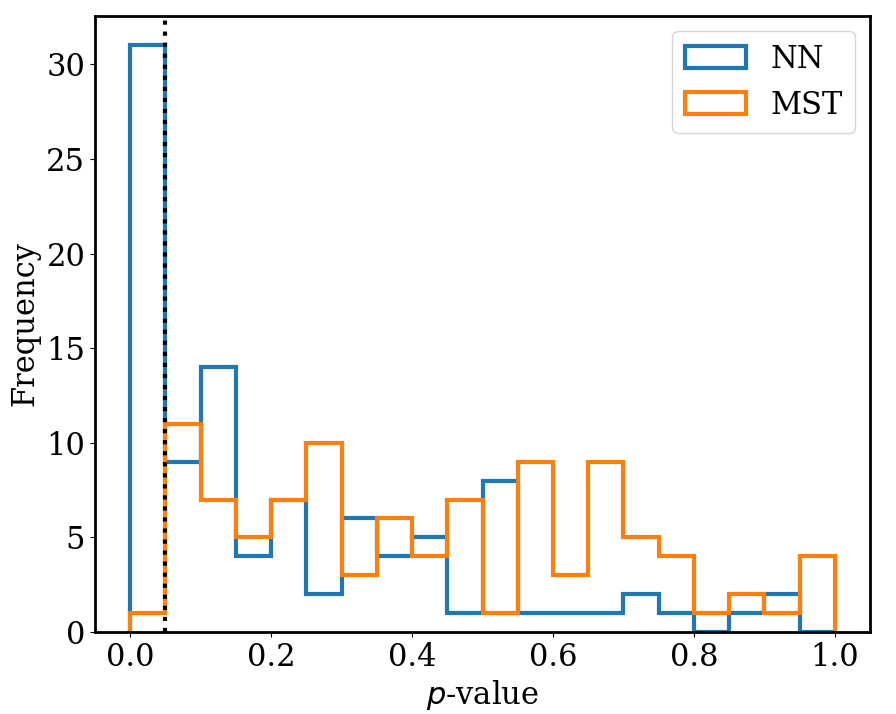}
\caption{As figure \ref{fig::NHT_hist} but for p-values derived from the Kolmogorov-Smirnov test.}
\label{fig::KS_hist}
\end{figure*}

A null hypothesis test is a common technique for measuring the statistical significance of a result. The first step is to construct a `no-effect' hypothesis with the aim of finding it false, termed the null hypothesis. One cannot prove that a data set comes from a certain distribution, only the likelihood of it coming from that distribution. Thus, the likelihood of the data set coming from the null hypothesis is calculated and converted into a $p$-value. A $p$-value is a measure of the likelihood that an outcome as, or more, extreme as the data set would result from the null hypothesis. As such, the $p$-value lies between 0 and 1. The $p$-value is compared against a certain threshold, $\alpha$, and if found to be greater than the threshold the null hypothesis cannot be rejected, if found to be less than the threshold then the null hypothesis can be rejected. Note that, if the null hypothesis is not rejected, it does not mean it is correct, one only has the power to reject. This is why the hypothesis that is tested against is constructed with the aim of finding it false. A typical value of the threshold is $\alpha = 0.05$. This is the threshold used in this work. 

Here the null hypothesis is that the cores in the filament are randomly placed along its length, i.e. there is no correlation between core positions and no characteristic spacing. To populate the distribution produced assuming the null hypothesis is true, 10,000 realisations of $N$ randomly placed cores are created, where $N$ is the number of cores in the data set, and analysed using the same method as the data set. 

The nearest neighbour and minimum spanning tree techniques result in a distribution of separations which can be described by its median/mean and interquartile range/standard deviation, hereafter called the average and the width. For each of the 10,000 realisations of randomly placed cores the distribution average and width measurement is stored. These data are converted into a normalised 2D-PDF using a KDE. This 2D-PDF is the distribution of observed averages and widths if the null hypothesis is true. The average and width measurement of the data set can be placed in this 2D space and the probability of such an event given the null is calculated. 

Figure \ref{fig::2D_PDF} shows the 2D-PDF using the median/interquartile range and mean/standard deviation for both the nearest neighbour and the minimum spanning tree method. The nearest neighbour method (top row) produces PDFs which tend to small averages and widths due to its sensitivity to small separations. The minimum spanning tree PDFs (bottom row) have their most probable outcome at higher averages and widths than the nearest neighbour PDF. The mean/standard deviation PDFs are narrow in the mean but very wide in the standard deviation.

Figure \ref{fig::2D_PDF} also shows, as a red dot, the average and width from the single-tier fragmentation realisation discussed in subsections \ref{SSEC:RES:NNS} and \ref{SSEC:RES:MST}. It is clear that the average and width is far from the peak of the 2D-PDF for both measures of the average and width, and both methods. The white contour is set at the same probability as the average and width of the data set, all regions of the PDF outside the contour are less probable to occur. A $p$-value for the data set is constructed by integrating over the regions of the PDF outside this contour. For the nearest neighbour method the $p$-value for the median/interquartile range and mean/standard deviation analysis are 0.0005 and 0.0001 respectively. For the minimum spanning tree method the $p$-values are 0.0013 and 0.0016. In all cases the null can be rejected  as the $p$-value is below the threshold, $\alpha = 0.05$, and one can say that the cores are not randomly placed. Thus, the average and width measurements of the distribution can be used to estimate the underlying characteristic spacing.

The $p$-value for all 100 realisations of the single-tier fragmentation, using both methods, can be calculated. The left panel of figure \ref{fig::NHT_hist} shows the histogram of $p$-values for the single-tier fragmentation filaments. All $p$-values less than 0.05 allow the null hypothesis to be rejected. For the nearest neighbour method, the null can be rejected $61\%$ and $80\%$ of the time when using the median/interquartile range and mean/standard deviation respectively. For the minimum spanning tree, the null can be rejected $33\%$ and $88\%$ of realisations when using the median/interquartile range and mean/standard deviation respectively. The mean/standard deviation has greater power to reject the null in both methods, while the the nearest neighbour method is nearly twice as powerful as the minimum spanning tree method when using the median/interquartile range. Both methods can therefore detect a single characteristic fragmentation length-scale and are typically statistically significant even with few cores ($N\leq10$).

The same analysis can be applied to the two-tier fragmentation filaments; the right panel of figure \ref{fig::NHT_hist} shows the histogram of the $p$-values for all 100 realisations. It is clear that it is difficult to reject the null using either method. For the nearest neighbour method the median/interquartile range is better at rejecting the null but only has a success rate of $20\%$. For the minimum spanning tree method, the median/interquartile range is also better and also has a success rate of $20\%$. The average and width of the distribution of separations resulting from two-tier fragmentation are hard to distinguish from those of randomly place cores when using either method. This makes it difficult to obtain statistically significant signatures of the fragmentation with the low number statistics commonly encountered in observations.

\begin{figure*}
\includegraphics[width=0.48\linewidth]{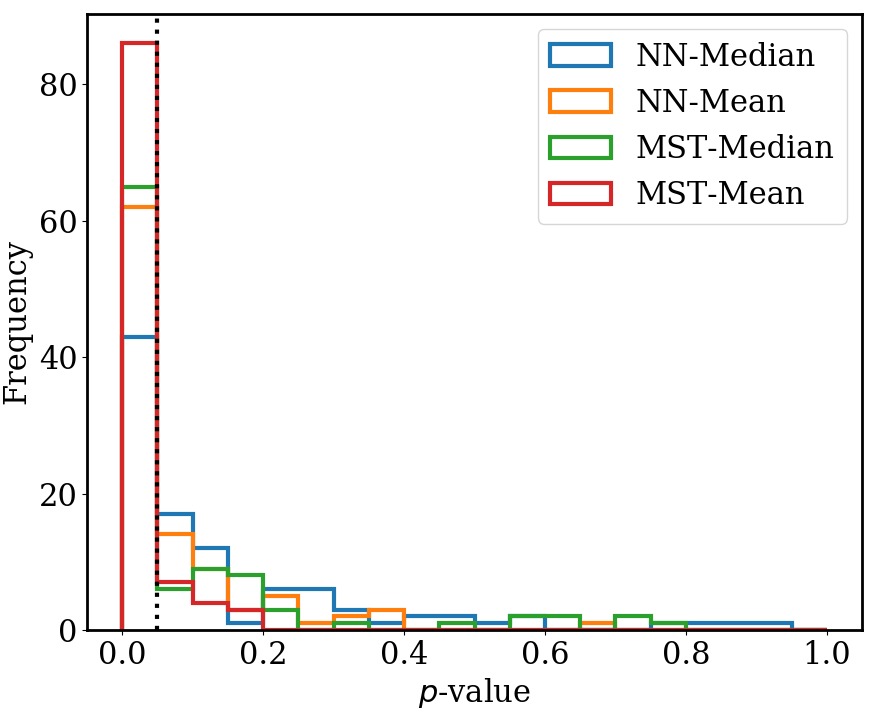}
\includegraphics[width=0.48\linewidth]{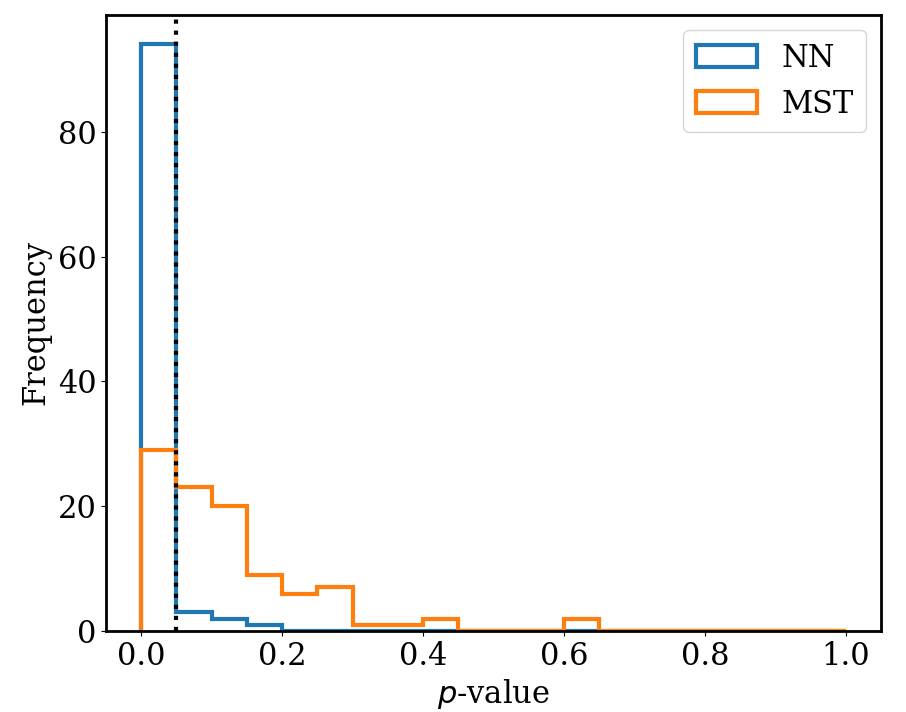}
\caption{Histograms of $p$-values (left) the null hypothesis test and (right) the KS-test for all 100 realisations with the fiducial `long' two-tier fragmentation parameters. The vertical dotted lines denotes the $p$-value threshold used to determine if the null hypothesis can be rejected, 0.05. In the legend, NN stands for nearest neighbour, and MST, for minimum spanning tree.}
\label{fig::long_NHT}
\end{figure*}

\subsubsection{Kolmogorov-Smirnov test}\label{SUBSUBSEC:KS}%

The Kolmogorov-Smirnov test (or the KS test) is a non-parametric test used to determine if two data-sets come from the same underlying distribution. As such, it is a null hypothesis test where the null hypothesis is that the two data-set do come from some underlying common distribution. The KS test determines the maximum distance between the cumulative distributions of the two data-sets, which can then be converted into a probability that such a distance occurs if the two data-sets arise from the same underlying distribution. 

For each of the 100 single-tier and two-tier fragmentation realisations, we run the KS-test on the separation distribution, either from the nearest neighbour or minimum spanning tree method, and compare to the separation distribution produced from 10,000 realisation of randomly placed cores, analysed using the same method. Figure \ref{fig::KS_hist} shows the histogram of $p$-values from the KS-test for both the single and two-tier fragmentation realisations.

For the single-tier fragmentation realisations (left), the KS-test is very powerful when determining the statistic significance of the results from the nearest neighbour method. The results from all 100 realisations are able to reject the null. It is less powerful for the minimum spanning tree method than the null hypothesis test presented in section \ref{SUBSUBSEC:NHT}; it is only able to reject the null hypothesis $51\%$ of the time. 

For the two-tier fragmentation, few realisations are able to reject the null for either the minimum spanning tree or nearest neighbour method. The null can be rejected $31\%$ of the time for the nearest neighbour method, slightly better than the null hypothesis test presented in section \ref{SUBSUBSEC:NHT}. However, the null can only be rejected $1\%$ of the time when using the minimum spanning tree method, significantly worse than the previously presented null hypothesis test.

We also investigate the use of the Anderson-Darling (AD) test, which is similar to the KS test but is more sensitive to deviations from normality. The results are not improved and are therefore not shown. The AD-test is also included in \textsc{FragMent}.

The results of the nearest neighbour and minimum spanning tree methods when applied to filaments which have fragmented in a two-tier manner are hard to distinguish from the results of randomly placed cores. This is true if one uses the KS(AD) test or the median/mean and interquartile range/standard deviation measurements. It is clear that these two methods are not sufficiently sensitive to multi-length-scale fragmentation to produce statistically robust results when there are only a few cores. This can be understood by the fact these methods only determine a single separation, either the nearest neighbour or the next core along the filament, and so require a large number of data points to adequately sample the underlying distribution.

\subsubsection{The effects of sample size}\label{SUBSUBSEC::N_NHT}%

A larger number of cores should help produce more statistically significant results. We produce a set of `long' filaments which are 8 pc long, and fragment in a two-tier manner using the fiducial parameters. These `long' filaments contain between 20 and 30 cores, roughly doubling the number of cores. 

Figure \ref{fig::long_NHT} shows the histograms of p-values from the null hypothesis test described above (\ref{SUBSUBSEC:NHT}), and the KS-test. For the null hypothesis test, the minimum spanning tree method using the mean/standard deviation is able to reject the null $86\%$ of the time, a large improvement over the normal two-tier fragmentation data-set where one can reject the null only $20\%$ of the time. For the KS-test, the nearest neighbour method is able to reject the null hypothesis $94\%$ of the time, considerably better than the $31\%$ of the time when there are only 10-15 cores. 

It appears that above $\sim 20$ cores one is able to reject the null most of the time if the fragmentation is more complicated than single-tier fragmentation. Below this number one may be able to reject the null, but it is significantly harder. 

\begin{table*}
\centering
\begin{tabular}{@{}*7l@{}}
\hline\hline
Realisations     & Mean NN & Median NN & Mean MST & Median MST & NN KS test & MST KS test  \\ \hline
Single-tier      & 80$\%$  & 61$\%$    & 88$\%$   & $33\%$     & 100$\%$    & 51$\%$       \\
Two-tier         & 2$\%$   & 20$\%$    & 7$\%$    & $20\%$     & 31$\%$     & 1$\%$        \\
`Long' two-tier  & 62$\%$  & 43$\%$    & 86$\%$   & $65\%$     & 94$\%$     & 29$\%$       \\ \hline \hline 
\end{tabular}
\centering
\caption{A table detailing the rate at which the null hypothesis can be rejected for the different fragmentation models. NN stands for nearest neighbour and MST stands for minimum spanning tree. }
\label{tab::NHT_res}
\end{table*}

\subsection{Single-tier vs two-tier fragmentation}\label{SSEC:Model}%

\begin{figure*}
\includegraphics[width=0.48\linewidth]{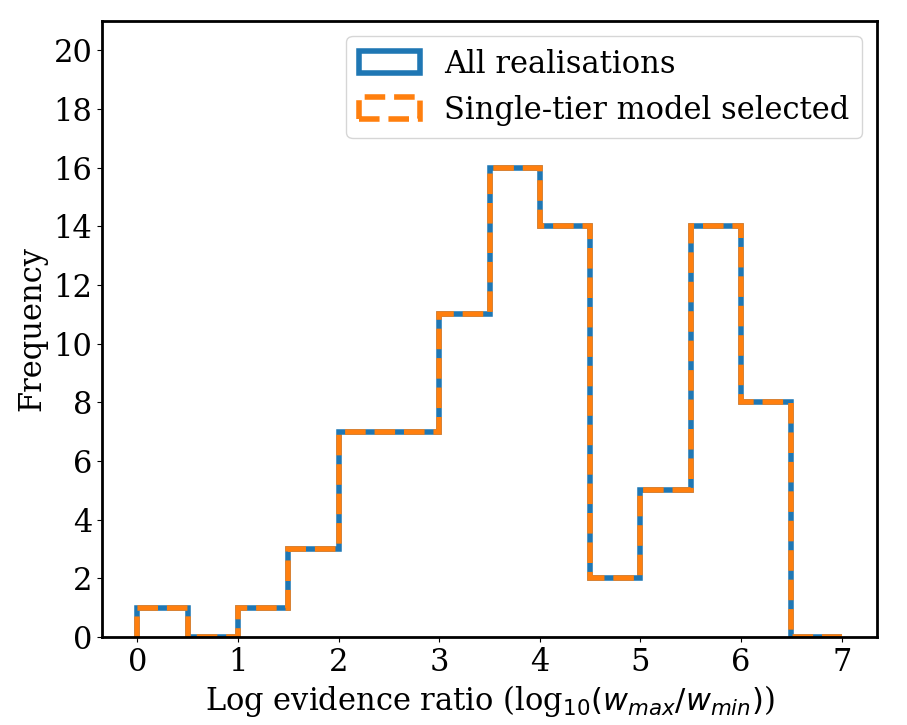}
\includegraphics[width=0.48\linewidth]{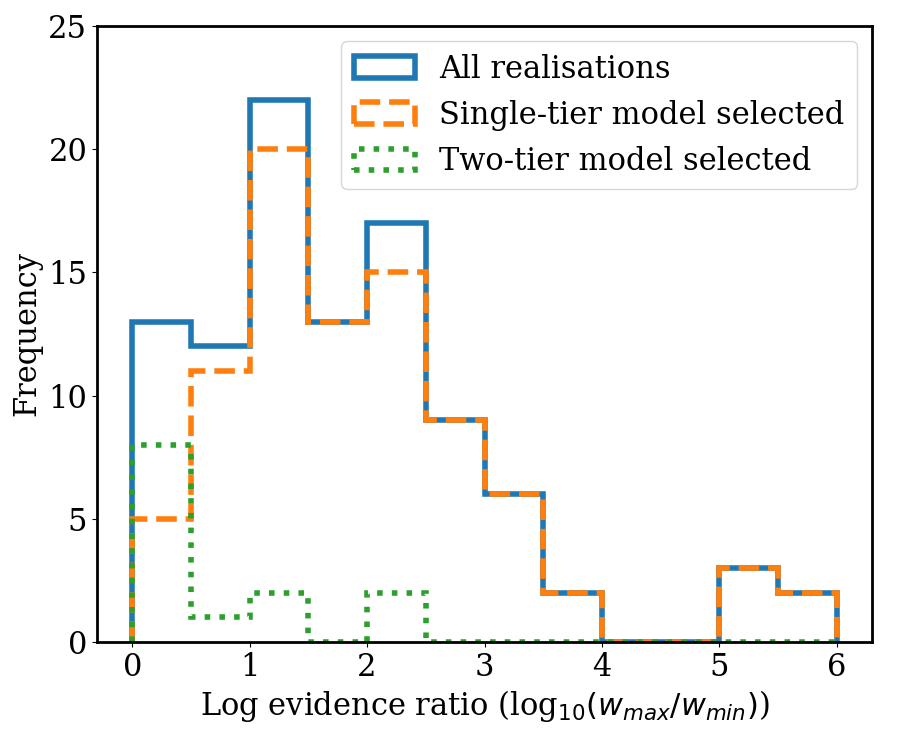}
\caption{Histograms of the log-evidence ratio for the (left) single-tier fragmentation realisations and (right) two-tier fragmentation realisations, both using the fiducial parameters. The blue line shows the evidence ratio for all realisations, the orange dashed line shows the same for realisations where the single-tier fragmentation model is selected as the best model, and the green dotted line shows the same for when the two-tier fragmentation model is selected.}
\label{fig::ak_weights}
\end{figure*}

\begin{figure*}
\includegraphics[width=0.48\linewidth]{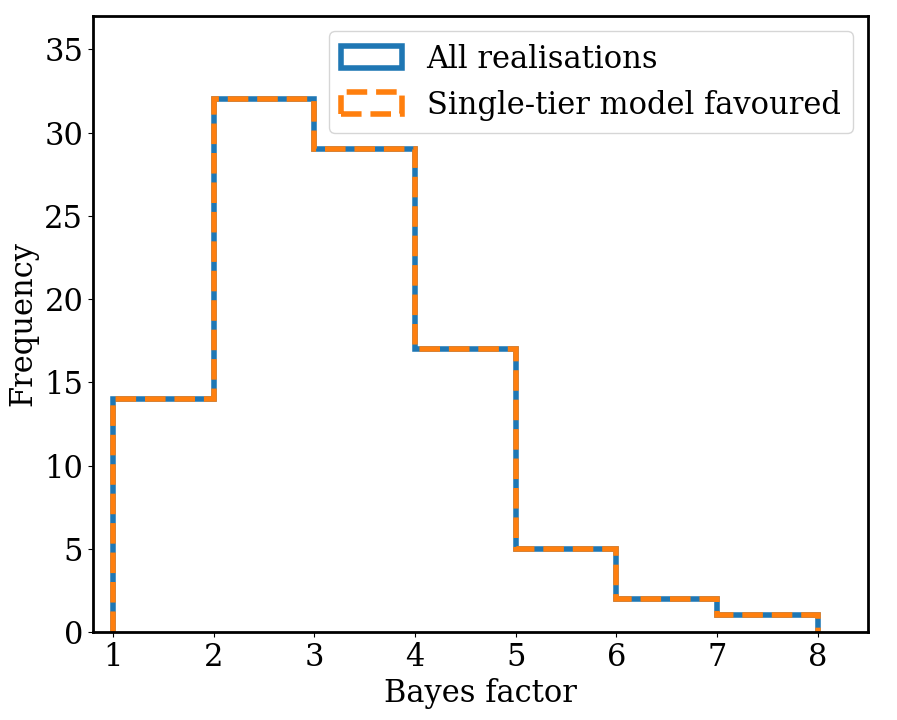}
\includegraphics[width=0.48\linewidth]{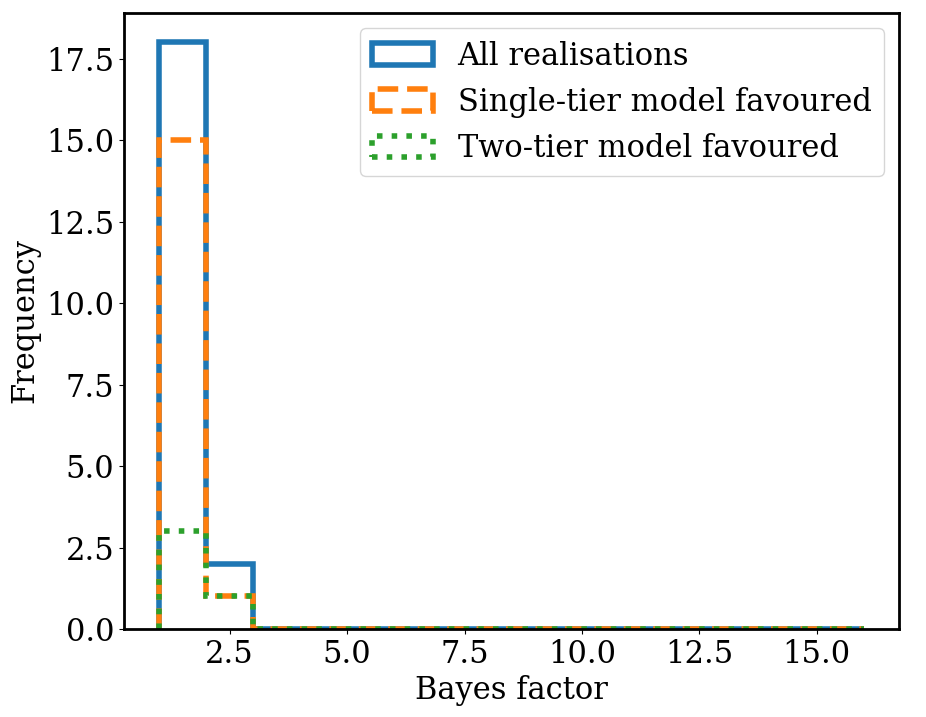}
\caption{Histograms of the Bayes factor, $\mathcal{B}$, for the (left) single-tier fragmentation realisations and (right) two-tier fragmentation realisations, both using the fiducial parameters. The blue line shows the Bayes factor for all realisations, the orange dashed line shows the same for realisations where the single-tier fragmentation model is favoured, and the green dotted line shows the same for when the two-tier fragmentation model is favoured. If the single- and two-tier fragmentation models are equally likely then $\mathcal{B} = 1$.}
\label{fig::Bayes-fac}
\end{figure*} 

A null hypothesis test can only provide the likelihood that the data-set is observed given a null hypothesis, here that there is no characteristic fragmentation length-scale. If the null hypothesis can be rejected, it is still unknown what type of non-random fragmentation exists. One may look at the results from the methods discussed and attempt to determine if the fragmentation is single-tier and can be described by a single characteristic fragmentation length-scale, or is two-tier and requires two such length-scales. Here we show two model selection methods included in \textsc{FragMent}: the frequentist approach using the Akaike information criterion, and the Bayesian approach using the odds ratio. 

\subsubsection{A frequentist model selection approach}\label{SUBSUBSEC:Freq}%

The underlying distribution that produced a data-set is unknown; therefore, one constructs a model to approximate this distribution. As the model is approximate it contains less information; the better the model approximates the underlying distribution, the less information is lost. The Akaike information criterion (AIC) is an estimator of this lost information and can therefore be used as a model selection tool \citep{Aka74}. 

The AIC of a model is defined as:
\begin{equation}
\mathrm{AIC} = 2k - 2\ln{L},
\end{equation} 
where $k$ is the number of parameters of the model and $L$ is the maximum likelihood calculated from the model. It is clear that AIC penalises models with a high number of parameters $k$; such models require a corresponding increase in the log-likelihood to offset its effect. For small sample sizes this penalisation is insufficient and the AIC can over-fit data. This small-sample size bias can be corrected; the modified information criterion is typically called the AIC$_c$, defined as:
\begin{equation}
\mathrm{AIC}_c = 2k - 2\ln{L} + \frac{2k^2 + 2k}{n - k - 1},
\end{equation} 
where $n$ is the sample size \citep{Sug78}. It is advisable to use the AIC$_c$ over the AIC when the ratio $n/k$ is greater than $\sim 40$ for the model with the highest number of parameters \citep{BurAnd02}. Note that as $n \to \infty$, AIC$_c$ $\to$ AIC; as such \textsc{FragMent} only determines the AIC$_c$. 

The exact value of AIC$_c$ is unimportant, it is the difference between the AIC$_c$ values of the models which is important. More precisely, using the differences one may construct the Akaike weights, which allows one to estimate the evidence of each model considered. The Akaike weight for model $i$ is defined as,
\begin{equation}
w_i = \frac{e^{-\frac{1}{2}\Delta_i}}{\sum_{j}^{N} e^{-\frac{1}{2}\Delta_j}},
\end{equation}
where $\Delta_i$ is the difference between the AIC$_c$ of model $i$ and the model with the minimum AIC$_c$, model $m$, and $N$ is the number of models being compared \citep{Aka78}. From this definition the model which is favoured, model $m$, has the minimum AIC$_c$ and the maximum Akaike weight. If the weight of model $i$ is 10 times lower than that of model $m$, then model $m$ is 10 times more likely and model $i$ can reasonably be rejected \citep{BurAnd02}.

There are two models to be examined here: that the fragmentation can be described by a single characteristic length-scale with some intrinsic scatter, or the fragmentation can be described by two characteristic length-scales, both with some scatter. These models are the same as the models used to construct the synthetic filaments. The single-tier model can be described by 2 parameters, the mean and standard deviation of a Gaussian. The data likelihood from this model is given by:
\begin{equation}
P(D|\underline{\theta}) = \prod_{i}^{M} \frac{1}{\sqrt{2\pi\sigma^2}} e^{-(x_i - \mu)^2/2 \sigma^2},
\end{equation}
where $x_i$ is the $i^{th}$ data point from a set of $M$, and $\underline{\theta}$ is the model parameter vector, ($\mu,\sigma$). The two-tier model can be described by 5 parameters, the mean and standard deviation of the two Gaussians and the ratio of the amplitude between the two, $\mathcal{R}$. The data likelihood from this model is given by:
\begin{equation}
P(D|\underline{\theta}) = \prod_{i}^{M} A_1 e^{-(x_i - \mu_1)^2/2 \sigma_{1}^2} + A_1 \mathcal{R} e^{-(x_i - \mu_2)^2/2 \sigma_{2}^2},
\end{equation}
where $\underline{\theta}$ is the model parameter vector, ($\mu_1,\sigma_1,\mu_2,\sigma_2,\mathcal{R}$), and $A_1 \equiv 1/(\sqrt{2\pi\sigma_1^2} + \mathcal{R}\sqrt{2\pi\sigma_2^2} )$.

Here, the data is the separation distribution resulting from applying the minimum spanning tree method. The minimum spanning tree method is used as it does not over-sample small spacings as the nearest neighbour approach does. Determining the data-likelihood, $P(D|\underline{\theta})$, while well sampling the parameter range can be computationally expensive. Moreover, it is only the maximum data likelihood which is needed for the calculation of the AIC$_c$. Therefore, \textsc{FragMent} uses a Monte-Carlo Markov-Chain routine \citep[\textsc{emcee},][]{For13} to efficiently sample the data-likelihood and find the maximum.

Figure \ref{fig::ak_weights} shows histograms of the evidence ratio (i.e. the weight of the `best' model, $w_{max}$, over the weight of the other model, $w_{min}$) for all 100 realisations of the fiducial parameters of both single-tier (left) and two-tier (right) fragmentation. For the single-tier fragmentation realisations, the single-tier model is selected as the best model $100\%$ of the time, and $99\%$ of the time the evidence ratio is greater than ten. It is therefore easy to determine the presence of single-tier fragmentation when it exists even with the small sample size here, $N_{core} \lesssim 10$. For the two-tier fragmentation realisations, the two-tier model is only selected 13$\%$ of the time, and only 4$\%$ of the time is the two-tier model selected and has an evidence ratio greater than 10. The single-tier model is selected 87$\%$ of the time, and $71\%$ of the time with an evidence ratio greater than 10. The increased number of parameters in the two-tier model often out-weighs the better data-likelihood found, and the two-tier realisations are mistaken for single-tier fragmentation. This is in line with the results from the null hypothesis testing in section \ref{SUBSEC:STATS}, which showed that it is difficult to reject the null hypothesis that the cores are randomly placed.  

\begin{table*}
\centering
\begin{tabular}{@{}*5l@{}}
\hline\hline
Frequentist (section \ref{SUBSUBSEC:Freq}) \\ \hline
Realisations     & Single-tier selected & Single-tier strongly & Two-tier selected & Two-tier strongly \\ 
 & & /positively selected & & /positively selected \\ \hline
Single-tier      & 100$\%$ (100$\%$)    & 99$\%$ ($99\%$)               & 0$\%$ (0$\%$)     & $0\%$ (0$\%$)  \\
Two-tier         & 87$\%$ (90$\%$)      & 71$\%$ (75$\%$)               & 13$\%$ (10$\%$)   & $4\%$ (0$\%$)  \\
`Long' two-tier  & 33$\%$ (38$\%$)      & 3$\%$  (3$\%$)                & 67$\%$ (62$\%$)   & $24\%$ (21$\%$)\\ \hline \hline 
Bayesian (section \ref{SUBSUBSEC:Bayes}) \\ \hline
Single-tier      & 100$\%$ (100$\%$)    & 54$\%$ (60$\%$)               & 0$\%$ (0$\%$)     & $0\%$ (0$\%$)  \\
Two-tier         & 75$\%$ (80$\%$)      & 4$\%$ (0$\%$)                 & 25$\%$ (20$\%$)   & $4\%$ (0$\%$)  \\
`Long' two-tier  & 61$\%$ (68$\%$)      & 4$\%$ (5$\%$)                 & 39$\%$ (32$\%$)   & $15\%$ (14$\%$)\\ \hline \hline 
\end{tabular}
\centering
\caption{A table detailing the rate at which each model is selected using the frequentist and Bayesian approaches. In brackets are the rates when only those realisations which can reject the null hypothesis are considered.}
\label{tab::select_results}
\end{table*}

\subsubsection{A Bayesian model selection approach}\label{SUBSUBSEC:Bayes}%

Model selection in the Bayesian framework is achieved using the odds ratio, $\mathcal{O}$, defined as:
\begin{equation}
\mathcal{O} = \frac{p(M_2 | D)}{p(M_1 | D)},
\end{equation}
between two models, $M_1$ and $M_2$, given data-set $D$ \citep{Ive14}. Using Bayes' rule this can be rewritten as,
\begin{equation}
\mathcal{O} = \frac{p(D|M_2) p(M_2)}{p(D|M_1) p(M_1)} = \mathcal{B} \; \frac{p(M_2)}{p(M_1)},
\end{equation}
where $\mathcal{B}$ is the Bayes factor between the two models. If there is no prior information favouring one model over the other then $p(M_2)/p(M_1) = 1$, and the odds ratio reduces to the Bayes factor. 

The term $p(D|M)$ is called the marginal likelihood of model $M$. It is often also called the evidence of model $M$. The marginal likelihood can be calculated as follows,
\begin{equation}
p(D|M) = \int p(D|M,\underline{\theta}) \, p(\underline{\theta}|M) \; d\underline{\theta},
\end{equation}
where $\underline{\theta}$ is the parameter space of the model. One may notice that the integrand is the un-normalised posterior, and thus the marginalised likelihood is the normalisation constant for the posterior.

For the two models under consideration here, single-tier and two-tier fragmentation, the data-likelihoods are given above in section \ref{SUBSUBSEC:Freq}. The priors used, $p(\underline{\theta}|M)$, are flat so as to contain as little information as possible, i.e. $p(\underline{\theta}|M) = C$. To normalise the priors and find the constant $C$, boundaries are placed on the values of each of the parameters and are given by:
\begin{align*}
0.0 \, \leq \mu_1 \leq 1.0, \\
0.0 \, \leq \sigma_1 \leq 0.5,\\
0.0 \, \leq \mu_2 \leq 1.0,\\
0.0 \, \leq \sigma_2 \leq 0.5,\\
0.0 \, \leq \mathcal{R} \leq 10.0.
\end{align*}

Currently \textsc{FragMent} calculates the posterior in a grid of $n^k$ evenly spaced evaluation points, where $n$ is a positive integer and $k$ is the number of parameters of the model. This is inefficient and may require a high $n$ for strongly peaked posteriors. For the realisations studied here we find that $n=30$ is sufficient such that when $n$ is doubled the normalisation constant changes by $< 1\%$. In the future a nested-sampling method \citep[such at the one used in \textsc{MultiNest},][]{Fer09} will be added to \textsc{FragMent}, allowing for a more efficient sampling of the posterior and faster calculation of the marginal likelihood. 

To interpret the odds ratio, \citet{KasRaf95} present a scale of evidence, modified from an earlier scale suggested by \citet{Jef61}. It is as follows:
\begin{align*}
& \mathcal{O} < 1 \; \; \; \; \; \; & \textrm{Model 1 is favoured over model 2,}\\
1 < \, & \mathcal{O} <  3 &\textrm{Model 2 is barely supported over model 1,}\\
3 < \, & \mathcal{O} < 20 & \textrm{Positive evidence for model 2,} \\
20 < \, & \mathcal{O} < 150 & \textrm{Strong evidence for model 2,} \\
150 < \, & \mathcal{O} & \textrm{Very strong evidence for model 2.}
\end{align*}

Figure \ref{fig::Bayes-fac} shows the odds ratio for all 100 realisations of the fiducial parameters of both single-tier and two-tier fragmentation. We assume both models are equally likely and so the odds ratio is equal to the Bayes factor. For the single-tier fragmentation realisations, the single-tier model is favoured 100$\%$ of the time. However, the Bayes factors are rather low, all below 8. $54\%$ have a Bayes factor above 3 and so can be interpreted as positive evidence for the single-tier model over the two-tier model. Here the frequentist approach is more powerful than the Bayesian one, as it is able to strongly support the single-tier model 99$\%$ of the time. 

For the two-tier fragmentation realisations, in $25\%$ of cases the two-tier model is favoured, and $4\%$ of the time, the Bayes factor is greater than 3 in favour of the two-tier model. This is better than the frequentist approach, which only selected the two-tier case $13\%$ of the time and favoured the single-tier model strongly $71\%$ of the time. Here there is only positive support for the single-tier model $4\%$ of the time. However, it is clear that the Bayesian approach has the same difficulty as the frequentist approach: it rarely decisively favours the correct two-tier model with such low number statistics. 

Table \ref{tab::select_results} summarises the rate at which each model is favoured using the two different approaches. In brackets it also shows the rate at which each model is favoured when one examines only the realisations which could reject the null hypothesis test; however, there is very little difference between this subset and the whole data set.

As well as returning the favoured model, the two approaches to model selection return either the best-fit model parameters (frequentist), or a model parameter posterior (Bayesian). If the model is strongly supported then one may use these to make inferences about the characteristic fragmentation length-scales, complementing the methods discussed in section \ref{SEC:RES}.

\begin{figure*}
\includegraphics[width=0.48\linewidth]{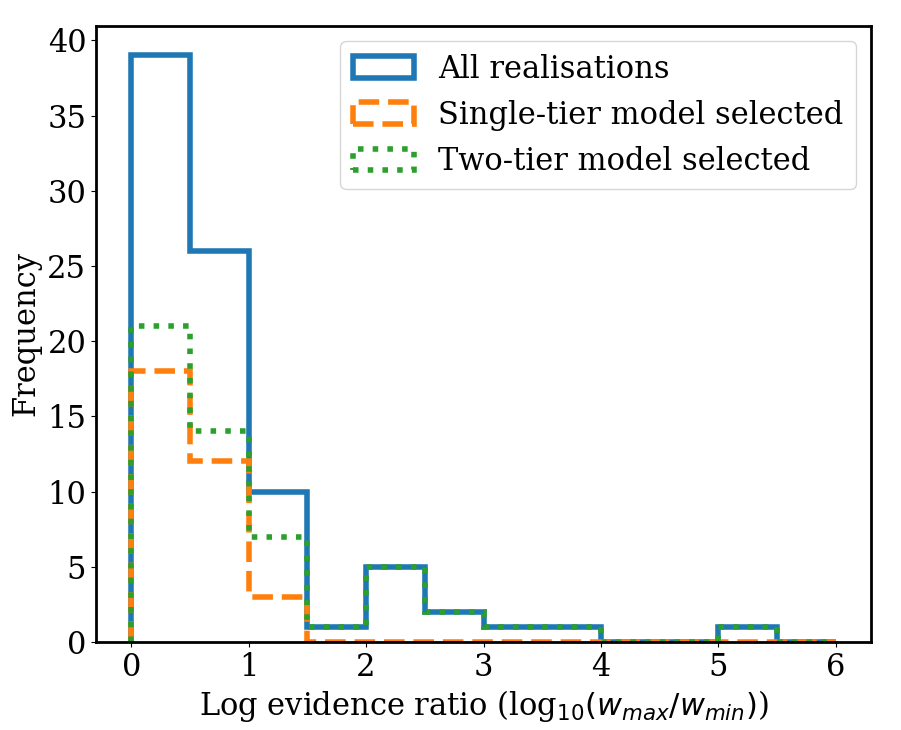}
\includegraphics[width=0.48\linewidth]{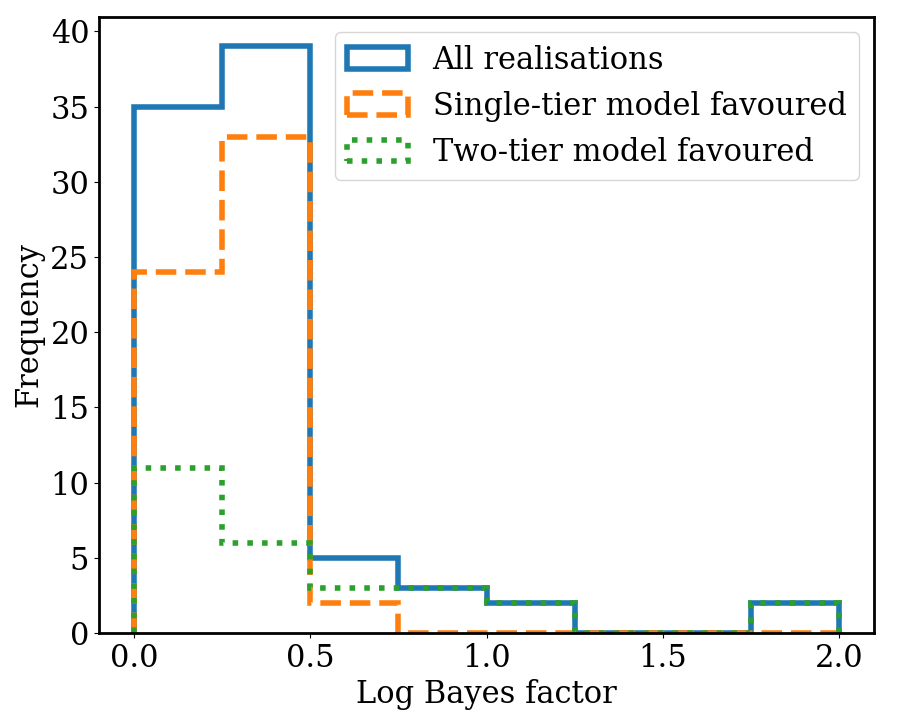}
\caption{Histograms of the logarithm of (left) the evidence ratio derived for the Akaike weights, and (right) the Bayes factor for the `long' two-tier fragmentation realisations using the fiducial parameters. The blue line shows the results for all realisations, the orange dashed line shows the same for realisations where the single-tier fragmentation model is favoured as the best model, and the green dotted line shows the same for when the two-tier fragmentation model is favoured.}
\label{fig::model_long}
\end{figure*} 

\subsubsection{The effects of sample size}%

For the small number of cores considered here ($N \lesssim 15$), both model selection techniques  erroneously, strongly favour the single-tier fragmentation model when the actual fragmentation pattern is two-tier. Here we consider the `long' (8 pc) two-tier fragmentation filaments introduced in section \ref{SUBSUBSEC::N_NHT}, which contain between 20 and 30 cores (roughly double the sample size). 

The left panel of figure \ref{fig::model_long} shows the histogram of the evidence ratio for the `long' two-tier fragmentation filaments using the fiducial parameters. Here the two-tier model is selected 67$\%$ of the time, and 24$\%$ of the time with an evidence ratio greater than 10. However, when the single-tier model is selected, it rarely has an evidence ratio greater than 10, only $3\%$ of the time. 

The right panel of figure \ref{fig::model_long} shows the histogram of the logarithm of the Bayes factors for all 100 `long' two-tier fragmentation realisations using the fiducial parameters. Here the maximum Bayes factor is $\sim 70$. The two-tier fragmentation model is favoured $39\%$ of the time, positively favoured $9\%$ of the time, and strongly favoured $6\%$ of the time. The single-tier model is only positively favoured $4\%$ of the time. This is slightly worse than the frequentist approach, which strongly supports the two-tier model $24\%$ of the time.    

With an increase in the number of cores the model selection improves in two ways: the single-tier is no longer favoured strongly the majority of the time, and the two-tier model becomes strongly favoured a sizeable minority of the time. It is difficult to decisively favour the correct model for two-tier fragmentation but when $N \gtrsim 20$ one can be confident that they are unlikely to strongly favour the incorrect model.

\section{Conclusions}\label{SEC:CON}%

Numerous theoretical works suggest that the fragmentation pattern of a filament contains information about the dynamical state of the filament, e.g. the relative importance of magnetic fields, gravity and turbulence. A number of techniques can be used, and have been used, in observational studies to determine characteristic fragmentation length-scales so that they may be compared to theory.  

The nearest neighbour separation, minimum spanning tree separation, and two-point correlation function are all sensitive to characteristic fragmentation length-scales and should be used together when investigating the fragmentation of a filament. The N$^{th}$ nearest neighbour and Fourier power spectrum are both poor techniques which are insensitive to characteristic fragmentation length-scales when there exists some variance in the value, as expected from real filaments. We discourage their use.

While some methods are sensitive to characteristic fragmentation length-scales, one must test the results for statistical significance. We devise a null hypothesis test, where the null hypothesis states that the cores are randomly placed along the filament, using the average of the separation distribution and its width. This is used to evaluate the results of the nearest neighbour and minimum spanning tree methods. The two-point correlation function includes a test for significance in its calculation. A Kolmogorov-Smirnov test is also used to test the results of the nearest neighbour and minimum spanning tree methods.   

When fragmentation is single-tier, the nearest neighbour, minimum spanning tree and two-point correlation function can routinely return statistically significant results, even if the number of cores is low, $N \lesssim 10$. When fragmentation is two-tier, the two-point correlation function is the only method to routinely produce statistically significant results for low number statistics ($N \lesssim 15$). However, when the number statistics improve ($N \gtrsim 20$), the nearest neighbour and minimum spanning tree methods can also return significant results. The nearest neighbour approach is typically better at rejecting the null when one uses the KS test, while the minimum spanning tree method is better at rejecting the null when comparing the average and width of the separation distribution against randomly placed cores. 

If the null is rejected, there is the question of how many characteristic fragmentation length-scales best describe the fragmentation. This can be addressed with the frequentist and Bayesian model selection techniques to determine if the fragmentation is best described by single-tier fragmentation or two-tier fragmentation. This should be applied to the minimum spanning tree separation distribution as it better represents the underlying separations, unlike the nearest neighbour separation distribution, which over-samples small separations by definition. The frequentist approach using the Akaike information criterion is typically better at selecting the correct model, and favouring it strongly. As with the null hypothesis test, it is difficult to select the two-tier fragmentation model when the fragmentation is two-tier. Typically at least 20 cores are needed to correctly favour the two-tier model. 

The fragmentation techniques, statistical significance tests, and model selection approaches shown here are provided to the community as an open-source \textsc{Python/C} library called \textsc{FragMent} at the URL \textit{https://github.com/SeamusClarke/FragMent}. 

\section{Acknowledgments}\label{SEC:ACK}%
The authors would like to thank the anonymous referee for their helpful comments on the paper. SDC and SW acknowledge support from the ERC starting grant No. 679852 `RADFEEDBACK'. GMW acknowledges support from the UK's Science and Technology Facilities Council under grant number ST/R000905/1. JCIM and SW thank the DFG for funding through the Collaborative Research Center (SFB956) on the `Conditions and impact of star formation' (sub-project C5).

\bibliographystyle{mn2e}
\bibliography{ref} 

\appendix

\section{FragMent: a Python/C package to study filament fragmentation}\label{APP:FRAG}%

\textsc{FragMent} includes the fragmentation techniques, statistical significance tests, and model selection approaches detailed in this paper, as well as an algorithm for straightening filaments. While the aim of this work is to study filament fragmentation, \textsc{FragMent} is written in such a way that all functions can be applied to any selection of 2D points. \textsc{FragMent} is open source and can be found at \textit{https://github.com/SeamusClarke/FragMent}

Here we detail 2 novel techniques that are included in \textsc{FragMent}: the method by which one may straighten a filament, and a tree-based algorithm to produce an approximate KDE. 

\subsection{Straightening filaments}\label{app::straight}%

\begin{figure*}
\includegraphics[width = 0.95\linewidth]{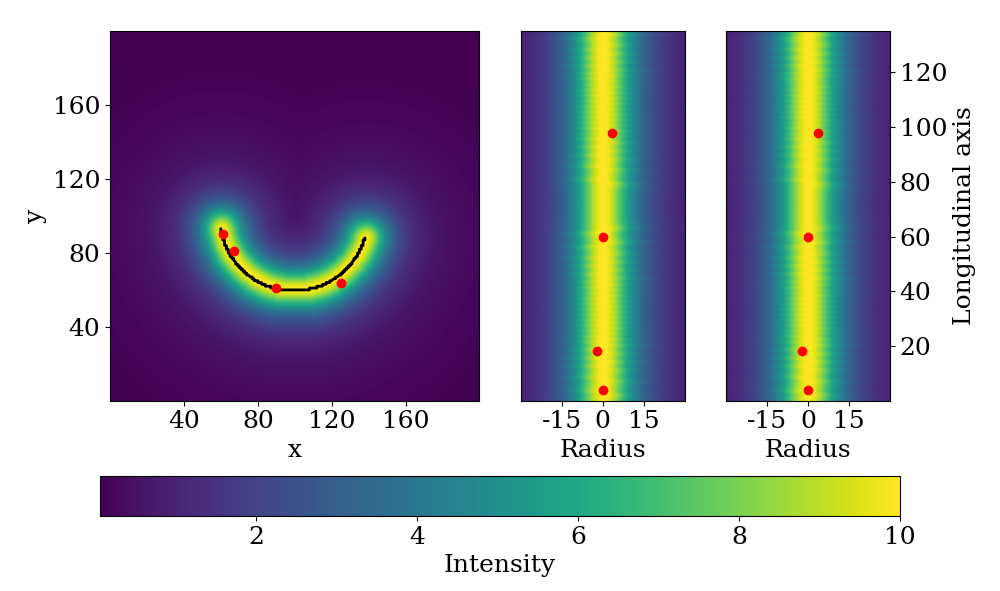}
\caption{(Left) A synthetic `intensity' map showing a curved filament. The solid black line shows the spine of the filament. (Middle) The straightened filament resulting from using the interpolation method to determine the radial profile. (Right) The straightened filament resulting from using the weighted average method to determine the radial profile.}
\label{fig::app:curve_fil}
\end{figure*}

In real filaments there exists curvature in the spine, resulting in the local radial axis of the filament not being perpendicular to the global longitudinal axis of the filament. To straighten a filament means to align the filament spine with the $y$-axis and for the radial profile at all points to lie parallel to the $x$-axis. This is seen in an example synthetic filament in figure \ref{fig::app:curve_fil}.

To straighten a filament, \textsc{FragMent} requires the column density / integrated intensity map and the list of spine pixel co-ordinates. The list of spine co-ordinates must be ordered such that the next physical spine point follows the current, i.e. the spine points go from one end to the other in a continuous manner. There are two methods for determining the radial profile: interpolation and a weighted average. Interpolation is used by \citet{Wil18} when they produce their straighten filaments. Both methods are included in \textsc{FragMent} and shown in \ref{fig::app:curve_fil}. The red dots shown on \ref{fig::app:curve_fil} represent `core positions' and how they appear when mapped onto the straightened filament using the \textsc{FragMent} function \textsc{Map$\_$Cores}.

For both methods, the first step is to fit a $n^{th}$ order polynominal to the spine points. This is done so that an accurate tangent can be found at every spine point. The tangent is the local longitudinal axis of the filament and thus the line perpendicular to it is the radial axis. When the value along the radial axis is determined by interpolation, we use the Taylor expansion up to and including the second order terms. When the value along the radial axis is determined by a weighted average of nearby pixels, the weighting function is a Gaussian:
\begin{equation}
W = \frac{1}{\sqrt{2 \pi h^2}} e^{-d^2 / 2 h^2},
\end{equation}
where $h$ is a smoothing-length and $d$ is the distance between the pixel and the evaluation point. 

For the interpolation method, there are 3 user-defined parameters: the distance to which the radial profile is measured, \textsc{max$\_$dist}; the number of evaluation points along the radial profile, \textsc{npix}; and the order of the polynominal used to fit the spine points, $n$. For the weighted average method, an extra parameter is used, the smoothing length of the weighting function, $h$. Here we have used \textsc{max$\_$dist} = 30, \textsc{npix} = 60, $n=10$, and $h=0.5$. The profiles only weakly depend on $n$ due to the simple filament shape and symmetry in the filament. A value of $h \gtrsim 1$ leads to over-smoothing at the filament's peak.   

\begin{figure*}
\includegraphics[width = 0.48\linewidth]{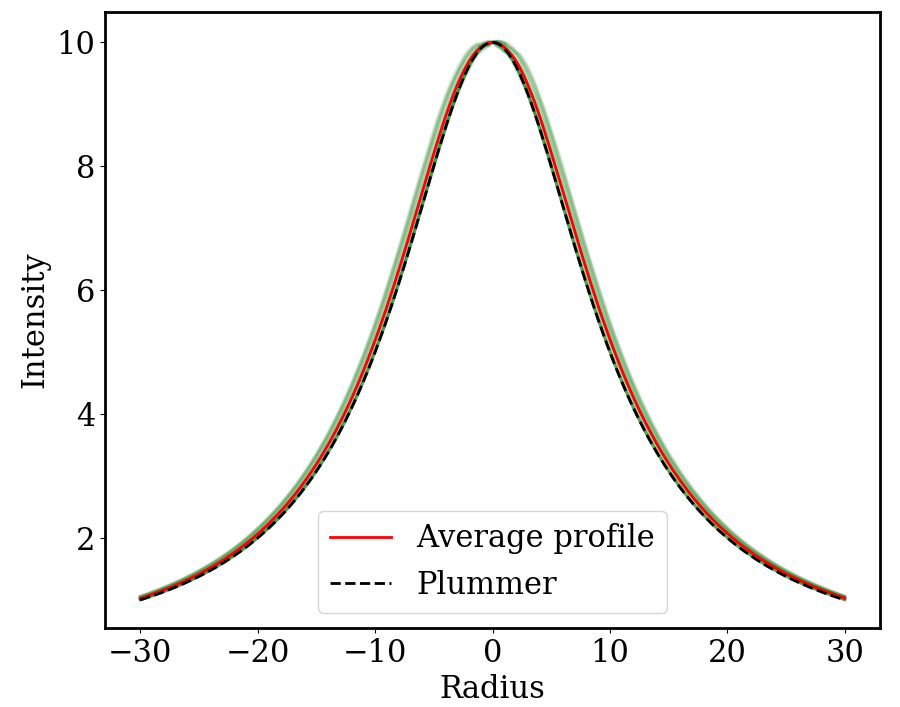}
\includegraphics[width = 0.48\linewidth]{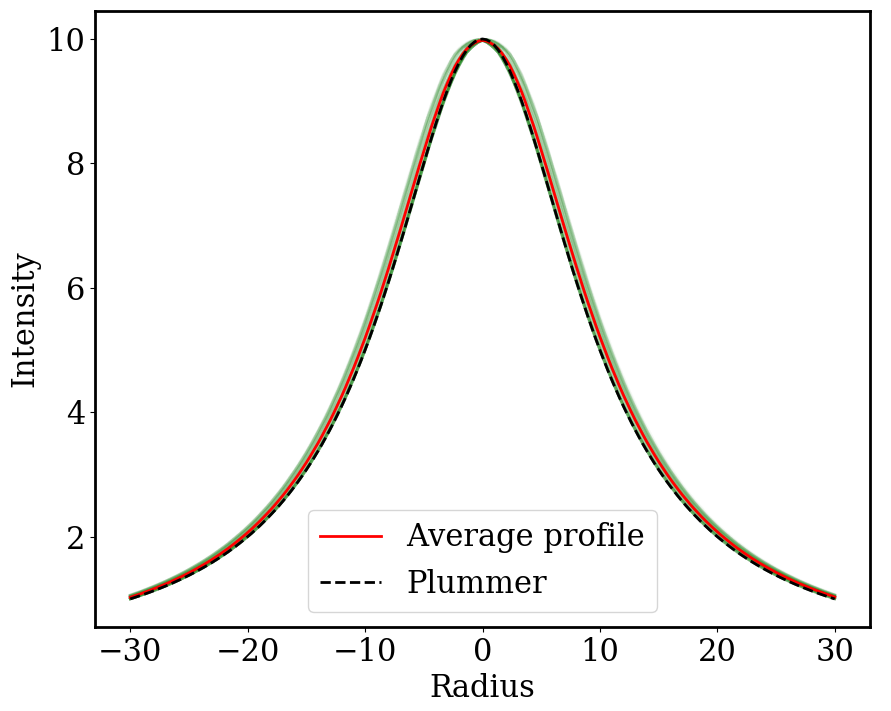}
\caption{The radial profile determine by (left) the interpolation method and (right) the weighted average method. In faded green is shown each individual radial profile, the red solid line shows the average radial profile, and the dashed black line shows the Plummer profile used to construct the synthetic filament.}
\label{fig::app:rad}
\end{figure*}

Figure \ref{fig::app:curve_fil} shows that the difference between these two methods is slight. Figure \ref{fig::app:rad} shows the radial profiles given by both methods. The two methods result in nearly identical radial profiles, with both methods recovering the Plummer profile used to construct the synthetic filament very well.

The stripes seen in both straightened filaments are artefacts which comes about when the tangent changes significantly. When the tangent changes significantly, the radial profile intersects different pixels leading to a lack of correlation between neighbouring radial profiles. This artefact is enhanced in the synthetic filament presented here as it is an arc of a circle, therefore the tangent is significantly changing from spine point to spine point. 

\subsection{A method for producing fast approximate Kernel Density Estimators}\label{APP:FRAG:AKDE}%

\begin{figure}
\includegraphics[width = 0.95\linewidth]{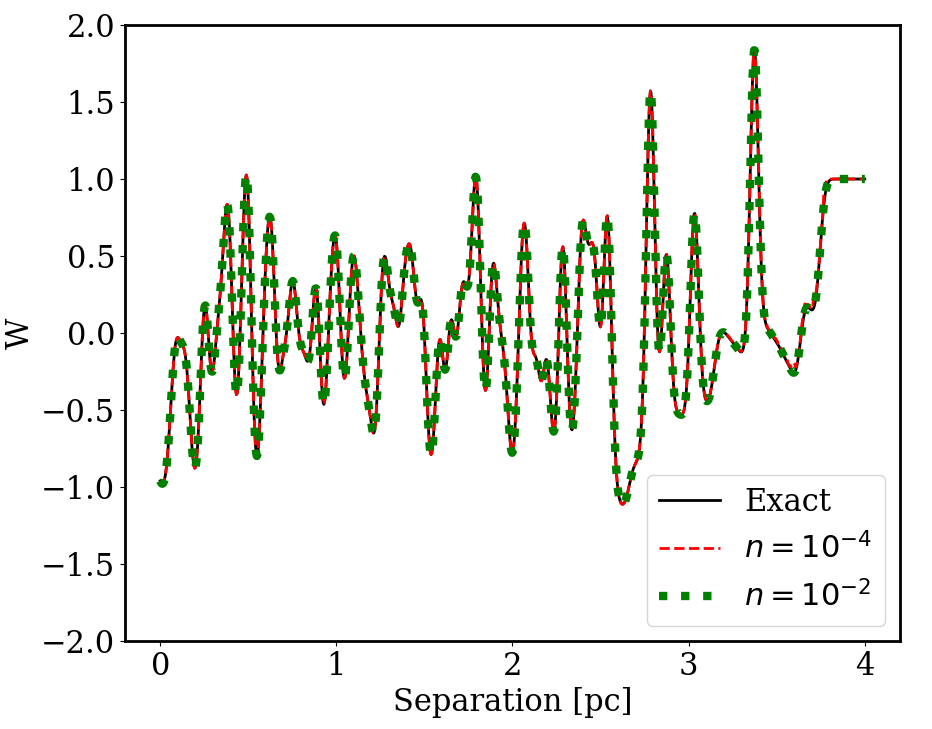}
\caption{The exact and two approximate KDEs of the same two-point correlation function. The 3 lines are near indistinguishable.}
\label{fig::akde}
\end{figure}

For large data-sets producing an exact KDE can be computationally expensive as each data point is convolved with the kernel. For a KDE constructed from $N_d$ data points, the evaluation of $N_e$ points scales as $\mathcal{O}(N_e N_d)$. When constructing the two-point correlation function described in section \ref{SSEC:RES:2P} the number of data points in the $DR$ array is $N_{core}^2 N_{real}$, where $N_{core}$ is the number of cores and $N_{real}$ is the number of realisations of randomly placed cores. As $N_{core} \sim 15$ and $N_{real} = 100,000$, the number of data points is $\sim 22,000,000$. Here we describe a tree-based method for producing fast approximate Kernel Density Estimators. 

Tree structures are commonly used in modern hydrodynamic codes to efficiently solve for the gravitational force. This can be done because when the source is far away, its contribution to the total gravitational acceleration is small. The same is true of a KDE as the kernel (here a Gaussian) tends to zero far from the data point. Thus, data points far away from the evaluation point can be treated as a group instead of individually, this group being a node in the tree. The evaluation of such an approximate KDE scales as $\mathcal{O}(N_e \log{N_d})$.

The tree structure used here is a binary tree. This means that the whole domain is considered as the root node in the tree which is then split into two child nodes. A node is split in this manner repeatedly until the resulting child nodes contain at most a certain number of data points, $N_{max}$. Nodes which do not have children are called leaf nodes. The threshold used here is 8 data points, i.e. all leaf nodes have at most 8 data points in them. 

Each node contains 4 pieces of information:
\begin{itemize}
\item the number of data points within the node,
\item the `data-centre' of the node,
\item the node boundaries,
\item and pointers to its two child nodes.
\end{itemize}
The `data-centre' of the node is the mean location of all data points in the node. 

\begin{table*}
\centering
\begin{tabular}{@{}*7l@{}}
\hline\hline
$n$ & $N_{max}$ & $N_{d}$ & median error & maximum error & time taken (sec) & speed-up factor                \\ \hline
Exact      & n/a & 22,500,000 & n/a                          & n/a                          & 377.785 & 1.00    \\ \hline

$10^{-4}$  & 4   & 22,500,000 & $7.69 \times 10^{-2} \; \%$  & 10.33 $\%$                   & 5.903   & 64.00   \\
$10^{-3}$  & 4   & 22,500,000 & $7.46 \times 10^{-4} \; \%$  & $3.90 \times 10^{-2} \; \%$  & 6.179   & 61.13   \\
$10^{-2}$  & 4   & 22,500,000 & $7.66 \times 10^{-6} \; \%$  & $3.58 \times 10^{-4} \; \%$  & 8.427   & 44.81   \\
$10^{-1}$  & 4   & 22,500,000 & $7.70 \times 10^{-8} \; \%$  & $4.09 \times 10^{-6} \; \%$  & 24.062  & 15.70   \\
1          & 4   & 22,500,000 & $2.34 \times 10^{-10} \; \%$ & $3.52 \times 10^{-10} \; \%$ & 101.334 & 3.73    \\ \hline

$10^{-3}$  & 8   & 22,500,000 & $7.46 \times 10^{-4} \; \%$  & $3.90 \times 10^{-2} \; \%$  & 5.667   & 67.00   \\
$10^{-3}$  & 16  & 22,500,000 & $7.46 \times 10^{-4} \; \%$  & $3.90 \times 10^{-2} \; \%$  & 5.509   & 68.56   \\ \hline 

$10^{-3}$  & 8   & 2,250,000  & $4.99 \times 10^{-2} \; \%$  & 2.98 $\%$                    & 0.495   & 103.18  \\
$10^{-2}$  & 8   & 2,250,000  & $4.77 \times 10^{-4} \; \%$  & $1.66 \times 10^{-2} \; \%$  & 0.697   & 73.75  \\
$10^{-3}$  & 8   & 225,000    & 2.57 $\%$                    & 152.30 $\%$                  & 0.044   & 133.76  \\ 
$10^{-2}$  & 8   & 225,000    & $2.92 \times 10^{-2} \; \%$  & 0.94 $\%$                    & 0.074   & 78.77  \\ \hline\hline

\end{tabular}
\centering
\caption{A table detailing the scaling and errors of the approximate KDE algorithm for various values of $n$, $N_{max}$ and $N_d$.}
\label{tab::akde}
\end{table*}

To evaluate the KDE at a point, the tree is `walked'. Starting from the root node, it is determined if the contribution from the node can be approximated as a single data point located at the node's centre and weighted with the number of data points the node contains. If the node contribution cannot be approximated it is said to be `opened' and the contribution from its child nodes (or the data points themselves if the node is a leaf node) is considered. To determine if the node is to be opened, one of two things must be true: the evaluation point is inside the node's boundaries, or the node is closer than some `opening' distance. By setting this opening distance one can control the error in the approximation, the larger the opening distance the closer the calculation becomes to the exact calculation. Here we use an opening distance of the form $n \sigma A$, where $n$ is a numerical constant, $\sigma$ is the bandwidth of the KDE, and $A$ is the number of data points in the node.

\subsubsection{Controlling the error}

By varying the numerical constant $n$ the error is controlled. Figure \ref{fig::akde} shows the exact KDE, and 2 approximate KDEs where $n \, = \, 10^{-4}$ and $10^{-2}$. The data shown is the same two-point correlation function shown in the right panel of figure \ref{fig::2Point}, only now showing the whole range of separations. The approximate KDEs are near indistinguishable from the exact KDE, the only small differences being at large separations.

Table \ref{tab::akde} shows the median error and maximum error in percent for a number of values of $n$. Table \ref{tab::akde} also shows the time taken for the calculation and the speed-up factor with respect to the exact solution. The time shown is not the time taken to compute the two-point correlation as this involved 3 KDEs ($DD, DR, RR$) and the computation of all the data points in these 3 arrays. Rather the time shown is the time taken to compute the KDE for the $DR$ array. With $n=10^{-3}$, the approximate KDE can achieves a maximum error of less than 0.1$\%$ with a factor of $\sim 60$ speed up. Increasing the maximum number of data points allowed in a leaf cell, $N_{max}$ does not affect the accuracy of the approximate KDE but does increase the speed up factor slightly. As a default $N_{max}=8$.
  
For fewer data points, $N_{d}$, the speed up factor is larger for the same value of $n$ but the error also increases. The results for 2,250,000 and 225,000 data points are shown in table \ref{tab::akde}. A general rule is that to achieve less than a 0.1$\%$ maximum error $n\geq 10^{4} / N_{d}$. 

The scaling with $N_d$ shown in table \ref{tab::akde} appears to be $\mathcal{O}(N_d)$ despite the fact walking a tree goes as $\mathcal{O}(\log{N_d})$. This is because for small values of $n$ the construction of the tree dominates the time of the approximate KDE algorithm, $\sim 90 \%$, and this scales as $\mathcal{O}(N_d)$. For larger $n$, the approximate KDE algorithm does scale as $\mathcal{O}(\log{N_d})$ as expected.

\label{lastpage}

\end{document}